\documentclass[aps,prb,reprint,superscriptaddress,footinbib,nobalancelastpage,amsmath,amssymb]{revtex4-1}

\usepackage{graphicx}% Include figure files
\usepackage{amssymb} %Contains math symbols
\usepackage{amsmath} %Contains math symbols
\usepackage{amsfonts}
\usepackage{slashed} %Feyman slash notation
\usepackage{comment}

\usepackage[english]{babel}
\usepackage{lmodern} %improves font style
\usepackage{bm} %bold math symbols (can only be used in math environment \bm{#}
\usepackage{hyperref}
\usepackage{xcolor}
\usepackage[nolist]{acronym} %nolist implies that no list of acronyms is created
\usepackage{braket} %bra ket notation
\usepackage[T1]{fontenc} 
\usepackage[normalem]{ulem}

\newcommand{\ii}      {\mathrm{i}}
\newcommand{\ee}      {\mathrm{e}}
\newcommand{\abs}[1]  {\left\lvert#1\right\rvert}

\newcommand{\diag}[1]   {\mathrm{Diag} \left[ #1 \right] }
\newcommand{\sgn}[1]        {\mathrm{sgn}\!\left(#1\right)}
\newcommand{\commutator}[1]   {\left[#1 \right]}

\newcommand{\unitVecX}  {\mathbf{e}_x}
\newcommand{\kvec}      {\mathbf{k}}
\newcommand{\xvec}      {\mathbf{x}}
\newcommand{\etaH}      {\eta(\Bz)}
\newcommand{\etaHR}      {\eta_\kappa^{}(\Bz)}
\newcommand{\etaSpin}[1]      {\eta^{}_{#1}\left(\Bz\right)}
\newcommand{\etaVac}    {\eta(0)}
\newcommand{\etaVacR}    {\eta_\kappa^{} (0)}
\newcommand{\myInt}[1]  {\int d #1 \,}

\newcommand{\nm}        {\,\mathrm{nm}}
\newcommand{\meV}       {\,\mathrm{meV}}
\newcommand{\meVnm}     {\,\mathrm{meVnm}}

\newcommand{\tesla}     {\,\mathrm{T}}
\newcommand{\mTesla}    {\,\mathrm{mT}}

\newcommand{\Bz}        {B_\perp}
\newcommand{\spinup}    {\mathord\uparrow}
\newcommand{\spindown}    {\mathord\downarrow}
\newcommand{\lb}        {\left(}
\newcommand{\rb}        {\right)}

\begin{document}

\date{\today}
\title {Fate of Quantum Anomalous Hall Effect in the Presence of External Magnetic Fields and Particle-Hole Asymmetry}

\author{Jan~B\"{o}ttcher}
\affiliation{Institut f\"ur theoretische Physik (TP4) and W\"urzburg-Dresden Cluster of Excellence ct.qmat,
Universit\"at W\"{u}rzburg, Am Hubland, 97074 W\"urzburg, Germany}

\author{Christian~Tutschku}
\affiliation{Institut f\"ur theoretische Physik (TP4) and W\"urzburg-Dresden Cluster of Excellence ct.qmat,
Universit\"at W\"{u}rzburg, Am Hubland, 97074 W\"urzburg, Germany}

\author{Ewelina~M.~Hankiewicz}
\email{Ewelina.Hankiewicz@physik.uni-wuerzburg.de}
\affiliation{Institut f\"ur theoretische Physik (TP4) and W\"urzburg-Dresden Cluster of Excellence ct.qmat,
Universit\"at W\"{u}rzburg, Am Hubland, 97074 W\"urzburg, Germany}

%Guideline Max 500 words
\begin{abstract}
The quantum anomalous Hall (QAH) effect, a condensed matter analog of the parity anomaly, is characterized by a quantized Hall conductivity in the absence of an external magnetic field. 
However, it has been recently shown  that, even in the presence of Landau levels, the QAH effect can be distinguished from the conventional quantum Hall (QH) effect due to the parity anomaly. 
As a signature of this effect, we predicted coexistent counterpropagating  QAH and QH edge states. 
In the present work, we generalize these findings to QAH insulators with broken particle-hole symmetry. 
In particular, we derive the connection to the spectral asymmetry, which is a topological quantity arising in the context of Dirac-like systems. 
Moreover, it is shown that, depending on the magnetic field direction, particle-hole asymmetry strengthens or weakens the hybridization of the counterpropagating QH and QAH edge states. 
Implications for ferro- or paramagnetic topological insulators are derived which paves the way towards identifying signatures of the QAH effect even in external magnetic fields.
\end{abstract}

\maketitle%

%%%%%%%%%%%%%%%%%%%%%%%%%%%%%%%%%%%%%%%%%%%
%%%%		ACRONYM
%%%%%%%%%%%%%%%%%%%%%%%%%%%%%%%%%%%%%%%%%%%
%Use \ac{Kuerzel} to refer to acronam in singular. First time also long form is shown
%Use \acp{Kuerzel} to show acronym in plural
\begin{acronym}
	\acro{QH}{quantum Hall}
	\acro{QAH}{quantum
	anomalous Hall}
	\acro{QSH}{quantum spin Hall}
	\acro{TI}{topological insulator}
	\acro{LL}{Landau level}
	\acro{CS}{Chern-Simons}
	\acro{BHZ}{Bernevig-Hughes-Zhang}
	\acro{2D}{two-dimensional}
	\acro{PH}{particle-hole}
	\acro{TR}{time-reversal}
	\acro{QED}{quantum electrodynamics}
\end{acronym}

%%%%%%%%%%%%%%%%%%%%%%%%%%%%%%%%%%%%%%%%%%%%%%%%%%%%%%%%%%%%%%%%%%%%%%%%%%%%%%%%%%%%%%%
%%%%%%%%%%%%%%%%%%%%%%%%%%%%%%%%%%%%%%%%%%%%%%%%%%%%%%%%%%%%%%%%%%%%%%%%%%%%%%%%%%%%%%%
%%%%%%%%%%%%%%%%%%%%%%%%%%%%%%%%%%%%%%%%%%%%%%%%%%%%%%%%%%%%%%%%%%%%%%%%%%%%%%%%%%%%%%%
%%%%%%%%%%%%%%			INTRODUCTION
%%%%%%%%%%%%%%%%%%%%%%%%%%%%%%%%%%%%%%%%%%%%%%%%%%%%%%%%%%%%%%%%%%%%%%%%%%%%%%%%%%%%%%%
%%%%%%%%%%%%%%%%%%%%%%%%%%%%%%%%%%%%%%%%%%%%%%%%%%%%%%%%%%%%%%%%%%%%%%%%%%%%%%%%%%%%%%%
%%%%%%%%%%%%%%%%%%%%%%%%%%%%%%%%%%%%%%%%%%%%%%%%%%%%%%%%%%%%%%%%%%%%%%%%%%%%%%%%%%%%%%%

Breaking of parity and \ac{TR} symmetry is the underlying principle for chiral, dissipationless edge transport at the boundary of \ac{2D} systems \cite{Schnyder08,Chiu16}, manifesting itself in a quantized Hall conductivity $\sigma_{xy}$.  There are two prominent examples of this in 2D:
The \ac{QH} and the \ac{QAH} effect. In the first effect, parity and \ac{TR} symmetry are broken by an external out-of-plane magnetic field. As such, the \ac{QH} phase relies on the formation of \acp{LL}. In the second effect, these symmetries are violated by the band structure itself, even in the absence of a magnetic field \cite{Liu08}. More precisely, the \ac{QAH} phase is based upon the inversion of electron and hole subbands for a single spin direction which can be experimentally realized by the introduction of magnetic dopants  into topological insulators \cite{Liu08,Yu10,Chang13,Checkelsky14,Liu16}. In the context of high energy physics, this phenomenon, i.e., the requirement of a quantized $\sigma_{xy}$ in the absence of \acp{LL}, is known as \textit{parity anomaly} \cite{Niemi83,Redlich84,Redlich84PRL,Haldane88,Mulligan13,Witten16}. More generally,  the parity anomaly describes the necessity of a broken parity and \ac{TR} symmetry in odd spacetime dimensions for an odd number of Dirac fermions \cite{Redlich84,Redlich84PRL}. 

Albeit QH and QAH phases have different physical origins, they fall into the same symmetry class \footnote{This statement only holds in the absence of PH symmetry.} and are both described by a $\mathbb{Z}$-topological invariant \cite{Chiu16}. When a \ac{QAH} insulator is subjected to an external magnetic field, it is hence natural to ask whether unique signatures of the \ac{QAH} phase persist in the presence of \acp{LL}. With respect to actual solid-state materials, this issue is of particular importance for paramagnetic topological insulators, such as (Hg,Mn)Te, as in these materials  a finite magnetic field is required  to realize the \ac{QAH} phase \cite{Liu08,Liu16}.

Recently, this fundamental question was addressed in Ref.~\cite{Boettcher19}  focusing mainly on \ac{PH} symmetric Chern insulators. Therein, it was shown that the QAH topology remains encoded in the presence of a magnetic field in the bulk \ac{LL} spectrum by means of a particular topological quantity, the spectral asymmetry $\eta$ \cite{Atiyah75,Nakahara03}, which is a signature of the parity anomaly  \cite{Niemi84,Boyanovsky86}. The spectral asymmetry represents the difference in the amount of states between valence and conduction band. Furthermore, it was shown that QH and QAH edge states can coexist at a certain chemical potential.  In contrast to conventional QH phases, QH and QAH edge states can exhibit thereby different chiralities and are hence counterpropagating. They form a pair of helical-like edge states \cite{Wang13}.

In this work, we generalize the formalism developed in Ref.~\cite{Boettcher19} to generic \ac{QAH} insulators that exhibit a broken \ac{PH} symmetry. In this context, we show that a broken PH symmetry, contrary to the naive expectation, does not contribute to the spectral asymmetry. Instead, it acts as a magnetic-field-dependent chemical potential. Furthermore, we show that a broken \ac{PH} symmetry can either strengthen or weaken the hybridization of the coexisting, helical-like QH and QAH edge states,  depending on the magnetic field direction. Finally, we compare transport signatures of  para- and ferromagnetic topological insulators which are described by  the \ac{BHZ} model (consisting of two Chern insulators) \cite{Bernevig06,Koenig08}. For ferromagnetic topological insulators, we show that the key signature of the QAH phase in magnetic fields is a hysteresis-like behavior of the Hall conductivity. While this feature has been already confirmed experimentally \cite{Moodera15}, we show that it is limited to a regime, where the magnetization of the ferromagnet dominates over the external field. When the orbital part of the magnetic field starts to dominate, we predict a sudden drop of $\sigma_{xy}$ to zero. The experimental observation of this prediction is so far outstanding. In the case of paramagnetic topological insulators, we show that the Hall conductivity  follows a reentrant like behavior. This means  a transition with increasing magnetic fields from $0$ to $\pm e^2/h$ back to $0$. We show that both signatures are encoded in the spectral asymmetry and are as such a  representative of the parity anomaly.

The paper is structured as follows. We start in Sec.~\ref{sec:Model} by introducing  Chern insulators and the  \ac{BHZ} Hamiltonian.
In Sec.~\ref{sec:phsCI}, we briefly review and summarize how the spectral asymmetry manifests itself in the particle number of a \ac{PH} symmetric Chern insulator. Particularly, the connection to the parity anomaly is clarified.
In Sec.~\ref{sec:brokenPHS}, a generalization to Chern insulators with broken \ac{PH} symmetry is presented. A general expression for the Hall conductivity is derived in Sec.~\ref{sec:bulkHall}. 
Implications of a broken \ac{PH} symmetry on the coexistence of QH and \ac{QAH} edge states in magnetic fields are discussed in Secs.~\ref{sec:bulkboundary} and~\ref{sec:hybDiscussion}. All results are combined in Sec.~\ref{sec:BHZ} to discuss the role of the parity anomaly and the spectral asymmetry for the full \ac{BHZ} model. Differences between ferro- and paramagnetic topological insulators are explained, and experimental consequences are derived. We conclude in Sec.~\ref{sec:outlook} by providing a summary and an outlook. 

%%%%%%%%%%%%%%%%%%%%%%%%%%%%%%%%%%%%%%%%%%%%%%%%%%%%%%%%%%%%%%%%%%%%%%%%%%%%%%%%%%%%%%%
%%%%%%%%%%%%%%%%%%%%%%%%%%%%%%%%%%%%%%%%%%%%%%%%%%%%%%%%%%%%%%%%%%%%%%%%%%%%%%%%%%%%%%%
%%%%%%%%%%%%%%%%%%%%%%%%%%%%%%%%%%%%%%%%%%%%%%%%%%%%%%%%%%%%%%%%%%%%%%%%%%%%%%%%%%%%%%%
%%%%%%%%%%%%%%			MAIN TEXT
%%%%%%%%%%%%%%%%%%%%%%%%%%%%%%%%%%%%%%%%%%%%%%%%%%%%%%%%%%%%%%%%%%%%%%%%%%%%%%%%%%%%%%%
%%%%%%%%%%%%%%%%%%%%%%%%%%%%%%%%%%%%%%%%%%%%%%%%%%%%%%%%%%%%%%%%%%%%%%%%%%%%%%%%%%%%%%%
%%%%%%%%%%%%%%%%%%%%%%%%%%%%%%%%%%%%%%%%%%%%%%%%%%%%%%%%%%%%%%%%%%%%%%%%%%%%%%%%%%%%%%%

%%%%%%%%%%%%%%%%%%%%%%%%%%%%%%%%%%%%%%%
%%      MODEL 
%%%%%%%%%%%%%%%%%%%%%%%%%%%%%%%%%%%%%%%
\section{Model} \label{sec:Model}
Our starting point is the \ac{BHZ} model \cite{Bernevig06}
\begin{align}\label{eq:fullBHZ}
    H\lb \mathbf{k} \rb = \begin{pmatrix} h\lb \mathbf{k} \rb & 0 \\ 0 & h^*\lb \mathbf{-k} \rb \end{pmatrix} \, ,
\end{align}
where both spin blocks are connected by \ac{TR} and parity symmetry and each spin block is determined by a Chern insulator \footnote{We assume that the basis of the \ac{BHZ} model is given by $\{ \ket{a \spinup},\ket{b \uparrow}, \ket{a \spindown}, \ket{b \spindown} \}$, where $a$ and $b$ denote a pseudospin degree of freedom. We therefore refer to each block as a spin block while we refer to the internal degree of freedom as a pseudospin.}: 
\begin{align} \label{eq:chernInsulator}
    h\lb \mathbf{k} \rb = 
    \begin{pmatrix} 
        M-\lb B+D \rb k^2 & Ak_+ \\
        Ak_- & -M+\lb B-D \rb k^2
    \end{pmatrix} \, .
\end{align}
Here, $k^2=k_x^2+k_y^2$, $k_\pm = k_x \pm  \ii k_y$, $M$ is the Dirac mass, $B$ and $D$ are related to the effective mass, and $A$ couples the two subbands. In particular, the parameter $D$ breaks the \ac{PH} symmetry  \footnote{We  always assume that the bulk is insulating, i.e, $\abs{B} > \abs{D}$.}. For clarity, App.~B reviews the  symmetries of the BHZ model and compares them with those of a Chern insulator.

The bulk spectrum of the BHZ model is double degenerated and reads 
\begin{align}\label{eq:spectrumCI}
    E^\pm_s \lb \kvec \rb = -D k^2 \pm \sqrt{A^2 k^2 +\lb M - B k^2 \rb^2} \, ,
\end{align}
where $s= \{ \spinup,\spindown \}$ denotes either the spin up [$h(\kvec)$] or the spin down [$h^*(-\kvec)$] block.  For $M/B>0$, the system is in the \ac{QSH} phase and hosts a pair of counterpropagating, helical edge states in the bulk gap \cite{Bernevig06,Koenig08}.

Using a unitary transformation, we can recast Eq.~\eqref{eq:fullBHZ} into the following form,
\begin{align} \label{eq:unitaryTrafoBHZ}
    U H\lb \mathbf{k} \rb U^\dagger = \diag{h\lb\mathbf{k},M,B\rb \, , \, h\lb\mathbf{k},-M,-B\rb} \, ,
\end{align}
where $U=\diag{\sigma_0 \, , \,  \sigma_y}$. The dependence on the parameters $M$ and $B$ is written out explicitly to show that it is sufficient to obtain analytical results for a single spin block. Results for the second spin direction are then derived by replacing $M\rightarrow -M$ and $B\rightarrow-B$. 
Before we consider the case of broken \ac{PH} symmetry, we introduce first some general concepts and  briefly review the most important, recently derived, results for a single Chern insulator with $D=0$ \cite{Boettcher19}.

%%%%%%%%%%%%%%%%%%%%%%%%%%%%%%%%%%%%%%%
%% MODEL - CHERN INSULATOR WITH PHS 
%%%%%%%%%%%%%%%%%%%%%%%%%%%%%%%%%%%%%%%

\section{Spectral asymmetry: Chern insulator obeying PH symmetry} \label{sec:phsCI}

In contrast to the full \ac{BHZ} model, a Chern insulator [Eq.~\eqref{eq:chernInsulator}] is characterized by a broken \ac{TR} and parity symmetry by virtue of the Dirac mass $M$ and the non-relativistic mass $B$.
This comes along with an integer quantized Hall response in the bulk gap, even in the absence of a magnetic field. In particular,  $\sigma_{xy}=\nu \, e^2/h$, where the Chern number is given by \cite{Lu10}
\begin{align} \label{eq:intrinsicHall}
\nu=\frac{1}{2}\left[\sgn{M}+\sgn{B}\right] \,.
\end{align}
Hence, the system is described by a $\mathbb{Z}$-topological invariant \cite{Schnyder08}. We refer to a nontrivial Chern insulator ($M/B>0$)  synonymously as a \ac{QAH} insulator.

Note that the Hall conductivity does not vanish in the limit of $M,B\rightarrow0^\pm$, but depends on whether we approach the limit from above or below.  If we started  with a parity symmetric theory ($M=B=0$), we would need to introduce a parity-breaking regulator during the computation of $\sigma_{xy}$ \cite{Niemi83,Redlich84,Jackiw84}. This shows that a single, parity-invariant Chern insulator cannot exist in two (spatial) dimensions. This unique phenomenon is a peculiarity of odd spacetime dimensions and is known as the parity anomaly \cite{Redlich84}. The \textit{intrinsic} Chern number, given by Eq.~\eqref{eq:intrinsicHall} (defined by its unique relation to the parity-breaking mass terms $M$ and $B$), is therefore a  necessary consequence of the parity anomaly.  
A parity invariant theory is only possible in (2+1)D, if it is either  embedded within a higher dimensional space [in this case, (3+1)D] or by adding a second Chern insulator \cite{Redlich84,Mulligan13}. The latter scenario is relevant when we come back to the \ac{BHZ} model.

Let us now consider the effect of an external, out-of-plane magnetic field $\Bz$ which can be incorporated using the Peierls substitution in the Landau gauge $\mathbf{k} \rightarrow \mathbf{k} + e \mathbf{A} /\hbar$ with $\mathbf{A} =-y \Bz \unitVecX$.
We adopt periodic boundary conditions in the $x$-direction and require the wave functions to be square-integrable in the $y$-direction.  Analytic results for the corresponding bulk \ac{LL} energies are obtained by replacing the canonical momentum operators with ladder operators \cite{Koenig08}:
\begin{align}
    E_{n\neq0}^{\pm}&=-\frac{\sgn{e \Bz} B}{l_{\Bz}^2} \! \pm \! \sqrt{\frac{2 A^2 n}{l_{\Bz}^2} + \! \lb M-\frac{2 n B}{l_{\Bz}^2} \rb ^2},   \label{eq:nLLenergies} \\
    %%%%
    E_\mathrm{n=0}&=\sgn{e \Bz} \lb M-B/l_{\Bz}^2 \rb \, ,   \label{eq:0LLenergies} 
\end{align}
where $l_{\Bz}=\sqrt{\hbar / \abs{e \Bz}}$ is the magnetic length, $n$ is the \ac{LL} index, and we set $D=0$. The corresponding wave functions are for completeness given in App.~\ref{app:Wavefunctions}.

\begin{figure}[!t]
    \includegraphics[width=.8\columnwidth]{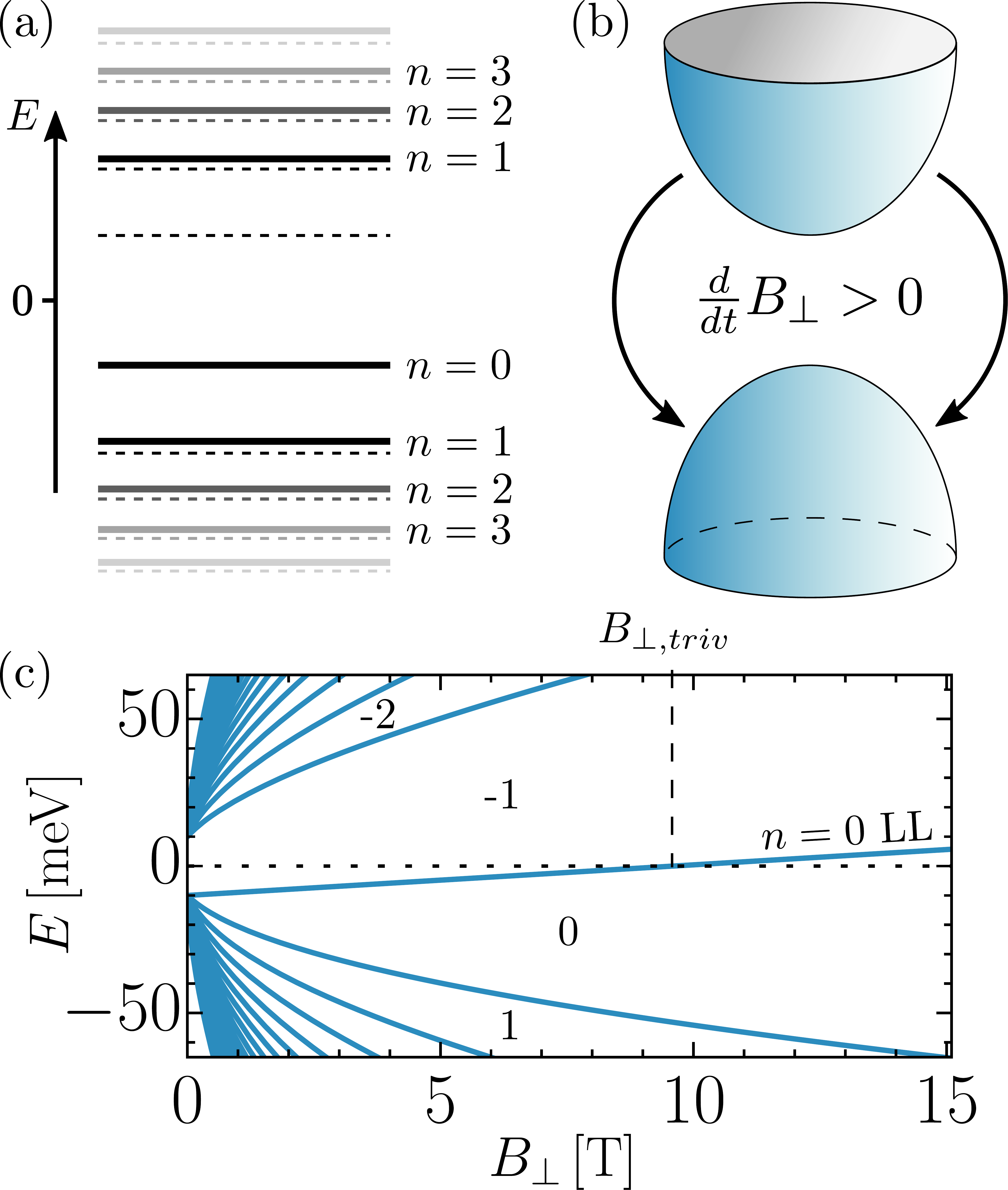}
    \caption{\label{fig:spectralAsymmetry} (a) Solid lines show sketch of LL spectrum for a Chern insulator. All \acp{LL} except for the $n=0$ \ac{LL} come in pairs. Mirroring the spectrum at $E=0$, depicted by dashed lines, highlights the asymmetry of the spectrum (introduced by $\Bz$). (b) Sketch indicates that with increasing magnetic field more and more states are removed from the conduction band and accumulate in the valence band provided that $M,B<0$. (c) Explicit evolution of bulk LL energies as function of $\Bz$ with $M=-10 \meV$, $B=-685 \meVnm^2$, $D=0$, and $A=365 \meVnm$. Dotted line indicates $E=0$. Dashed line marks $B_{\perp,triv}$, defined by Eq.~\eqref{eq:trivField}. Chern numbers are indicated.}
\end{figure}

The underlying \ac{PH} symmetry ($D=0$) is broken by applying the external magnetic field. This is on the one hand reflected by the formation of a single, unpaired $n=0$ \ac{LL} [see Figs.~\ref{fig:spectralAsymmetry}(a) and~(c)]. This specific \ac{LL} is located either in the valence ($E<0$) or in the conduction band ($E>0$) depending on $\sgn{e \Bz}\sgn{M-B/l_{\Bz}^2}$. On the other hand, although all  \acp{LL} with $n \geq 1$ come in pairs, they are not symmetric with respect to zero energy because of the non-relativistic mass parameter $B$. This is visualized by a sketch of the bulk LL energies in Fig.~\ref{fig:spectralAsymmetry}(a). To be precise, this additional contribution to the asymmetry is proportional to $\sgn{e \Bz}\sgn{B}$ [cf.~Eq.~\eqref{eq:nLLenergies}].

The corresponding bulk particle number in magnetic fields for an  arbitrary chemical potential $\mu$ can be calculated using the following expression  \cite{NiemiRev84}:
\begin{align}
\braket{\mathrel{N}}_{\mu,\Bz}&=\frac{1}{2} \myInt{\xvec} \sum_{\alpha=1}^2 \braket{\commutator{\Psi_\alpha^\dagger \lb \mathbf{x} \rb, \Psi_\alpha \lb \mathbf{x} \rb}}_{\mu,\Bz} \nonumber \\
&=\braket{N_0}_{\mu,\Bz}-\frac{\etaH}{2} \, . \label{eq:particleNumDef}
\end{align}
Here, $\braket{\cdots}$ marks the expectation value and $\Psi\lb\mathbf{x}\rb$ is the second-quantized fermionic field operator (two-component \ac{LL} spinor), whose explicit form is given in App.~\ref{app:Wavefunctions}. We use antisymmetrization as the appropriate operator ordering. For $D=0$, this choice is equivalent to normal ordering and ensures a vanishing particle number in the bulk gap at the charge neutrality point ($\mu=0$) for $\Bz=0$. Equation~\eqref{eq:particleNumDef} consists of two contributions: The first term denotes the expectation value of the `conventional' (fermionic) number operator $\braket{N_0}_{\mu,\Bz}$ with
\begin{align} \label{eq:bareFermionNumberOp}
N_0 = \sum_{n,k_x} b^\dagger_{n,k_x} b_{n,k_x}-\sum_{n,k_x} d^\dagger_{n,k_x} d_{n,k_x} \, , 
\end{align}
where $b_{n,k_x}$ destroys an electron in the $n$-th conduction band \ac{LL} with momentum $k_x$, and $d_{n,k_x}$ destroys a hole in the $n$-th valence band \ac{LL} with momentum $k_x$. Here, we mean  by `conventional' that $N_0$  consists only of fermionic operators and counts the number of filled and empty states with respect to the charge neutrality point at $\Bz=0$. In Eq.~\eqref{eq:bareFermionNumberOp}, the $n=0$ LL plays a special role, since it  belongs either to the first ($E_{n=0}>0$) or to the second ($E_{n=0}<0$) sum. This is because it is either part of the valence or  the conduction band, as we have pointed out before. The second term in Eq.~\eqref{eq:particleNumDef} is the spectral asymmetry  which is given by
\begin{align} \label{eq:spectralAsymDef}
\etaH = \sum_{E>0} 1 - \sum_{E<0} 1 = \sum_E \sgn{E} \, .
\end{align}
Roughly speaking, $\eta$  is a measure of the asymmetry of the entire spectrum. It is hence zero if the spectrum obeys a PH symmetry, 
\begin{align}
    U_C^\dagger h^\star[\kvec ; \mathbf{A}(\xvec)] U_C = -h[-\kvec ; \mathbf{A}(\xvec)] \, ,
\end{align}
or a chiral symmetry,
\begin{align}
    U_S^\dagger h[\kvec ; \mathbf{A}(\xvec)] U_S = -h[\kvec ; \mathbf{A}(\xvec)] \, ,
\end{align}%
%%%%%%%%%%%%%%%%%%%%%%%%%%%%
%%%%%%%%%%%%%%%%%%%%%%%%%%
\noindent
where $U_C$ and $U_S$ are unitary matrices [further details are given in App.~B]. This is because for every contribution in the first sum of Eq.~\eqref{eq:spectralAsymDef}, these symmetries guarantee exactly one corresponding term (with the opposite sign) in the second sum  \footnote{The statement holds only assuming that the system has no zero modes \cite{Witten16}}.
The spectral asymmetry can be only nonzero if PH and chiral symmetry are broken. It is for that reason no surprise that a magnetic field, which breaks the underlying \ac{PH} symmetry, can induce a nonzero $\eta$ in a \ac{QAH} insulator. More concretely, we showed explicitly in Ref.~\cite{Boettcher19} that, for $D=0$, 
\begin{align} \label{eq:spectralAsymPHS}
    \etaH = n_0 \, \sgn{e \Bz} \! \left[\sgn{\! M- \frac{B}{ l_{\Bz}^2}\!}\!+\sgn{B}\right]\!,  
\end{align}
where $n_0=S/\lb 2\pi l_{\Bz}^2 \rb$ is the \ac{LL} degeneracy and $S=L_x L_y$ is the area. This result consists of  two distinct contributions: Firstly, the asymmetry which arises from the existence of a single, unpaired $n=0$ \ac{LL}, reflected by $\sgn{e \Bz} \, \sgn{M-B/l_{\Bz}^2}$, and, secondly, from the asymmetry of all \acp{LL} with $n \geq 1$, reflected by $\sgn{e \Bz} \, \sgn{B}$. Comparing Eq.~\eqref{eq:intrinsicHall} with~\eqref{eq:spectralAsymPHS}, it is apparent that the spectral asymmetry is a direct consequence of the intrinsic Chern number and is as such a signature of the parity anomaly.

Furthermore, Eq.~\eqref{eq:spectralAsymPHS} shows that, for a QAH insulator with  $M,B<0$ and $\sgn{e \Bz}>0$, increasing  the magnetic field is accompanied by an increase in the spectral asymmetry. This means the magnetic field pushes successively more and more states from the conduction into the valence band. This is schematically depicted in Fig.~\ref{fig:spectralAsymmetry}(b). The amount of excess charge corresponds effectively to having in total one more \ac{LL} in the valence band. This process continues until the $n=0$ LL, which traverses the bulk gap for $B\neq 0$ [Fig.~\ref{fig:spectralAsymmetry}(c)], crosses over into the conduction band at
\begin{align} \label{eq:trivField}
B_{\perp,triv} = \sgn{e \Bz} \frac{\hbar M}{e B} \, .
\end{align}
This point is marked for clarity in Fig.~\ref{fig:spectralAsymmetry}(c).
Above this critical field, the two distinct contributions of $\eta$ cancel each other, implying that valence and conduction band contain the same number of states and $\eta = 0$. Equation~\eqref{eq:spectralAsymPHS} therefore links the special properties of the $n=0$ LL in a QAH insulator to a topological quantity, the spectral asymmetry $\etaH$. Note that, in contrast, a trivial Chern insulator exhibits $\eta=0$ independent of the magnetic field.

%%%%%%%%%%%%%%%%%%%%%%%%%%%%%%%%%%%%%%%%%%%%%%%%%%
%%     MODEL - CHERN INSULATOR WITHOUT PHS 
%%%%%%%%%%%%%%%%%%%%%%%%%%%%%%%%%%%%%%%%%%%%%%%%%%
\section{Spectral asymmetry: Chern insulator without PH symmetry} \label{sec:brokenPHS}

We now want to include the $D$ parameter   in our model, which breaks \ac{PH} symmetry already at $\Bz=0$. One might naively expect that introducing $D$ should alter the spectral asymmetry, Eq.~\eqref{eq:spectralAsymPHS}, as $\eta$ measures the asymmetry of the underlying spectrum. However, this would be in conflict with our statement that $\eta$ is a signature of the parity anomaly.  The $D$ parameter breaks neither parity nor TR symmetry and is from this point of view not expected to appear in the spectral asymmetry. 

To clarify this issue, let us first check whether antisymmetrization is still an appropriate way to derive the normal ordered particle number operator.  This means antisymmetrization should ensure a vanishing  particle number  in the ground state $\ket{\mathrm{vac}}$. To this end, we normal order all  fermionic operators  with respect to $E_z=-M D / B$, implying that $\braket{\mathrm{vac}|N_0|\mathrm{vac}}=0$, where $N_0$ is understood analogously to Eq.~\eqref{eq:bareFermionNumberOp}. This specific value is chosen because $E_z$ corresponds to the Dirac point and, hence, to the charge neutrality point of our system \cite{Zhou08}. To verify that the particle number vanishes in the bulk gap, we are left with computing the spectral asymmetry:
\begin{align} \label{eq:etaVac}
    \etaVac = \sum_{E > E_z} 1-\sum_{E < E_z} 1 \, .
\end{align}
Since both sums are separately divergent, a regularization procedure must be employed to evaluate this expression. Here, we choose a heat-kernel regulator \cite{Nakahara03}, meaning that in Eq.~\eqref{eq:etaVac} all summands must be replaced according to
\begin{align}
    E>E_z: \quad &  1 \rightarrow \ee^{-\kappa\left[ E^+ (\kvec)-E_z \right]} \, ,   \nonumber \\
    E<E_z: \quad &  1 \rightarrow \ee^{+\kappa\left[ E^- (\kvec)-E_z \right]} \, ,
\end{align}
where $E^\pm(\kvec)$ is given by Eq.~\eqref{eq:spectrumCI} (with $s=\spinup$). At the end of this calculation, the limit $\kappa\rightarrow 0^+$ must be performed to recover the original expression. The regularized spectral asymmetry takes the form:
\begin{align}\label{eq:etaVac1stStep}
    \etaVacR &= \sum_{\kvec} \ee^{-\kappa \left[ E^+(\kvec) - E_z\right]} - \sum_{\kvec} \ee^{\kappa \left[ E^-(\kvec)-E_z \right]} \, .
\end{align}
It is convenient to replace both sums by integrals in the continuum limit, i.e., $\sum_{k_x,k_y}\rightarrow S/(2\pi)^2 \myInt{k_x} \myInt{k_y}$. Additionally, we Taylor expand the eigenenergies in Eq.~\eqref{eq:etaVac1stStep} for large momenta to simplify the expression further [$\mathcal{O}(k^{-2})$]:
\begin{align} 
    E^\pm (\kvec) &= -D k^2 \pm \abs{B}k^2 \sqrt{1+\frac{A^2k^2-2B k^2M +M^2}{B^2 k^4}} \nonumber \\
    &\approx \pm \sgn{B}\lb \frac{A^2}{2B}-M \rb - \lb D \mp \abs{B} \rb k^2  \, . \label{eq:approximationNoBz}
\end{align}
Since the heat-kernel regulator in Eq.~\eqref{eq:etaVac1stStep} affects only large energy (momentum) solutions, this approximation becomes exact in the limit $\kappa \rightarrow 0^+$ [App.~\ref{app:spectralAsym}].
This allows us to recast Eq.~\eqref{eq:etaVac1stStep} into conventional Gaussian integrals,
\begin{multline} \label{eq:etaIntegralForm}
    \etaVacR=\frac{S}{(2\pi)^2} \ee^{-\kappa \, \sgn{B} \lb \frac{A^2}{2B}-M \rb } 
    \left[\ee^{\kappa E_z}\int_{\mathbb{R}^2_{}} d\kvec \, \ee^{-\kappa B_- k^2} \right. \\
    \left. -\ee^{-\kappa E_z}\int_{\mathbb{R}^2_{}} d \kvec \, \ee^{-\kappa B_+ k^2} \right] \, ,
\end{multline}
where $B_\pm = \abs{B} \pm D$. We arrive finally at the expression:
\begin{align} \label{eq:etaVacFinal}
    \etaVacR=\frac{S}{2\pi} \frac{D}{B^2-D^2}\lb \frac{1}{\kappa}-\frac{A^2}{2\abs{B}} \rb + \mathcal{O}(\kappa) \, .
\end{align}
Based on this result, it is clear that antisymmetrization is no longer equivalent to normal ordering. It does not to ensure a vanishing particle number in the bulk gap. 
This issue can be traced back to the \ac{PH} asymmetry, which breaks the one-to-one correspondence between terms in the first and the second sum of Eq.~\eqref{eq:spectralAsymDef}. As a result, even in the absence of a magnetic field, the  two divergent sums do not longer cancel each other. 
However based on Eq.~\eqref{eq:etaVacFinal}, we can now define a new, properly renormalized number operator $\tilde{N}$ by subtracting Eq.~\eqref{eq:etaVacFinal} from the antisymmetrized number operator at finite $\Bz$:
\begin{align}
    \braket{\tilde{N}}_{\mu,\Bz} & = 
    \braket{N}_{\mu,\Bz} - \braket{N}_{\mu=E_z,\Bz=0} \nonumber \\
    & = \braket{N_0}_{\mu,\Bz} - \lim_{\kappa \rightarrow 0^+}\left[\frac{\etaHR-\etaVacR}{2}\right] \, , \label{eq:renormParticle}
\end{align}
where $\eta_\kappa^{}(\Bz=0) = \etaVacR$. This redefinition is mandatory to fulfill the physical requirement that $\braket{\tilde{N}}_{\mu=E_z,\Bz=0}=0$ at the charge neutrality point. This choice is also consistent with Eq.~\eqref{eq:particleNumDef} in the limit $D=0$. 

To be finally in the position to evaluate the particle number for arbitrary $\mu$ and $\Bz$, we first have to calculate the spectral asymmetry $\etaHR$ [cf. Eq.~\eqref{eq:renormParticle}]. In that regard, we need the \ac{LL} energies for $D\neq0$  \cite{Koenig08}:
\begin{multline}
    E_{n\neq0}^{\pm}=-\frac{\sgn{e \Bz} B + 2 n D}{l_{\Bz}^2}   \pm \left[ \frac{2 A^2 n}{l_{\Bz}^2} \right. \\
    \left. +\lb M-\frac{2 n B+\sgn{e \Bz} D}{l_{\Bz}^2} \rb ^2\right]^{\frac{1}{2}} \label{eq:nLLenergiesD} \, ,
\end{multline}
and
\begin{align}
    E_\mathrm{n=0}&=\sgn{e \Bz} \lb M-\frac{B}{l_{\Bz}^2} \rb - \frac{D}{l_{\Bz}^2} \, .   \label{eq:LLenergiesD} 
\end{align}
Note that the $n=0$ LL crosses the charge neutrality point $E_z$ at $\Bz=B_{\perp,triv}$, which is the same critical magnetic field that we obtained for the \ac{PH} symmetric case, Eq.~\eqref{eq:trivField}.  Using again the heat-kernel regularization, the spectral asymmetry becomes
\begin{multline}\label{eq:spectralAsymSDFirst}
    \etaHR= s\sum_{k_x} \sgn{M-\frac{B}{l_{\Bz}^2}} \ee^{-\kappa \abs{E_0-E_z}}  \\
    + \sum_{k_x,n=1} \ee^{-\kappa \lb E_n^+ -E_z\rb }-\sum_{k_x,n=1} \ee^{\kappa \lb E_n^- - E_z\rb} \, ,
\end{multline}
where we introduced the abbreviation $s\equiv\sgn{e \Bz}$. The summation runs over all momenta $k_x$ and all \acp{LL}. The first term marks the contribution of the $n=0$ \ac{LL}. Recall that it is either part of the valence ($E<E_z$) or the conduction ($E>E_z$) band which is why it enters in Eq.~\eqref{eq:spectralAsymSDFirst} with $\sgn{e \Bz} \, \sgn{M-B/l_{\Bz}^2}$. The second and third term mark the contribution of all \acp{LL} with $n\geq1$. Analogously to the approximation used in Eq.~\eqref{eq:approximationNoBz}, we  Taylor expand the exponents in the latter equation for large $n$. The resulting approximation is exact in the limit $\kappa\rightarrow0^+$ [App.~\ref{app:spectralAsym}]. Equation~\eqref{eq:spectralAsymSDFirst} takes then the form:
\begin{widetext}
\begin{align}
    \etaHR =& \, n_0 \left\{ s \, \sgn{M-\frac{B}{l_{\Bz}^2}} \ee^{-\kappa \abs{E_0-E_z}} +
    \ee^{\kappa\, \sgn{B} \lb M-A^2/2B-s D/l_{\Bz}^2 \rb } \right. \nonumber \\
    &\left. \cdot\left[
    \ee^{\kappa\lb s B /l_{\Bz}^2+E_z\rb}\sum_{n=1}^\infty \ee^{-2 n \kappa \lb \abs{B}-D\rb / l_{\Bz}^2}-
    \ee^{-\kappa\lb s B /l_{\Bz}^2+E_z\rb}\sum_{n=1}^\infty \ee^{-2 n \kappa \lb \abs{B}+D\rb / l_{\Bz}^2}
    \right]\right\} \label{eq:longEq} \, ,
\end{align}
\end{widetext}
where $\sum_{k_x}=n_0$ is the \ac{LL} degeneracy. This result can be recast noting that both sums form geometric series. After additionally Taylor expanding this result for small $\kappa$, we find that Eq.~\eqref{eq:longEq} becomes
\begin{align} \label{eq:finalEta}
    \etaHR = \etaVacR+\etaH + \mathcal{O}(\kappa) \, ,
\end{align}
where $\etaVacR$ and $\etaH$ are given by Eq.~\eqref{eq:etaVacFinal} and  Eq.~\eqref{eq:spectralAsymPHS}, respectively. Inserting now Eq.~\eqref{eq:finalEta} into Eq.~\eqref{eq:renormParticle} and performing the limit $\kappa \rightarrow 0^+$,  we obtain the final result for the renormalized particle number in magnetic fields:
\begin{align} 
    \braket{\tilde{N}}_{\mu,\Bz} = \braket{N_0}_{\mu,\Bz} - \frac{\etaH}{2} \label{eq:finalCharge} \, ,
\end{align}
which matches exactly the result of the \ac{PH} symmetric case [Eq.~\eqref{eq:particleNumDef}]. The naive expectation that the $D$-parameter should contribute to the spectral asymmetry is wrong. Instead,  $\etaH$ depends only on the parity breaking mass terms $M$ and $B$. Physically, this underlines once more that the spectral asymmetry is a consequence of the parity anomaly and is therefore exclusively related to parity breaking mass terms. The explicit role of the $D$-parameter in magnetic fields will become clear in the following sections.

%%%%%%%%%%%%%%%%%%%%%%%%%%%%%%%%%%%%%%%%%%%%%%%%%%
%%     Hall conductivity - bulk perspective 
%%%%%%%%%%%%%%%%%%%%%%%%%%%%%%%%%%%%%%%%%%%%%%%%%%
\section{Hall conductivity: bulk perspective} \label{sec:bulkHall}
Having determined a general expression for the particle number in magnetic fields, the Hall conductivity $\sigma_{xy}$ can be computed using Streda's formula, $\sigma_{xy}=\partial \rho(\mu,\Bz) / \partial \Bz |_\mu$. Here, $\rho(\mu,\Bz)=-e\braket{\tilde{N}}_{\mu,\Bz}/S$ is the charge carrier density and $\braket{\tilde{N}}_{\mu,\Bz}$ is given by Eq.~\eqref{eq:finalCharge}. Based on the special form of the number operator, the Hall conductivity can be divided in two distinct contributions \footnote{The presented results for $\sigma_{xy}^{I/II}$  have been already presented by us in Ref.~\cite{Boettcher19}, but without proof for $D\neq0$.},
\begin{align} \label{eq:totalHall}
\sigma_{xy}(\mu,\Bz)=\sigma_{xy}^I(\Bz) + \sigma_{xy}^{II}(\mu,\Bz) \, . 
\end{align}
We define  the first term by its exclusive relation to the spectral asymmetry:
\begin{align}
    \sigma_{xy}^I  & = \frac{e}{2S}\frac{\partial \eta(\Bz)}{\partial \Bz} \nonumber \\
    & =\frac{e^2}{2 h} \left[ \sgn{M-\frac{B}{l_{\Bz}^2}} + \sgn{B} \right] \, . \label{eq:sigmaI}
\end{align}
This quantity is independent of the chemical potential as it is a property of the entire spectrum. In comparison, $\sigma_{xy}^{II}$ depends on $\mu$ since it comprises all contributions that are associated to the conventional number operator $N_0$:
\begin{align} \label{eq:sigmaIIDef}
    \sigma_{xy}^{II}=-\frac{e}{S}\frac{\partial}{\partial \Bz} \braket{\mu,\Bz | N_0 | \mu, \Bz } \, ,
\end{align}
where $\ket{\mu,\Bz}$ defines an arbitrary many-particle state. Recall that it is zero  in the ground state, since all fermionic operators are normal ordered with respect to $E_z$, meaning:
\begin{align} \label{eq:vacuumState}
    N_0 \ket{\mathrm{vac}}= N_0 \lb  \prod_{n=\{0,1\}}^\infty \prod_{k_x=-\infty}^\infty d_{n,k_x} \ket{0} \rb  =0 \, ,
\end{align}
where $n=\{0,1\}$ means that the product starts either at $n=0$ or $1$, depending on whether the $n=0$ LL is part of the valence or  the conduction band, respectively. Consequently, Eq.~\eqref{eq:sigmaIIDef} can be only nonzero for $\mu \neq E_z$, and is directly related to the number of filled, or empty, \acp{LL} with respect to $E_z$.

To find a general expression for $\sigma_{xy}^{II}(\mu,\Bz)$, we have to distinguish several cases due to the special properties of the $n=0$ \ac{LL}. Let us start with $M,B<0$ and $\abs{\Bz}<B_{\perp,triv}$, implying that the $n=0$ \ac{LL} is filled in the ground state. In that case, the many-particle state $\ket{\mu,\Bz}$ with $\mu<E_z$ and $\Bz>0$  reads
\begin{align}
    \ket{\mu,\Bz}=\prod_{n=0}^{N_{max}(\mu)} \prod_{k_x=-\infty}^{\infty}d_{n,k_x}^\dagger \ket{\mathrm{vac}} \, ,
\end{align}
where $N_{max}(\mu)$ gives the number of empty valence band \acp{LL}. Inserting this state into Eq.~\eqref{eq:sigmaIIDef}, we arrive at
\begin{align}
    \sigma_{xy}^{II}=\frac{e^2}{h} \left[\theta\lb-\mu+E_{n=0}\rb +\sum_{n=1}^\infty \theta\lb -\mu+E_{n}^- \rb \right] \, ,
\end{align}
where for the given magnetic field direction $E_{n=0}=M-(B+D)/l_{\Bz}^2$. Taking the same parameters but choosing $\mu > E_z$, the corresponding many-particle state reads
\begin{align}
    \ket{\mu,\Bz}=\prod_{n=1}^{N_{max}(\mu)} \prod_{k_x=-\infty}^{\infty}b_{n,k_x}^\dagger \ket{\mathrm{vac}} \, ,
\end{align}
where $N_{max}(\mu)$ is the number of filled conduction band \acp{LL}. The associated Hall conductivity is given by
\begin{align}
    \sigma_{xy}^{II}=-\frac{e^2}{h} \sum_{n=1}^\infty \theta\lb \mu-E_{n}^+ \rb \, .
\end{align}
The latter equation does not include the $n=0$ \ac{LL}, since it is already filled in the ground state for the given set of parameters.

In the next step, we keep the same parameters but flip the sign of the magnetic field, $\Bz<0$. Most importantly and in contrast to the case $\Bz>0$, the $n=0$ \ac{LL} is now unoccupied in the ground state. The arbitrary many-particle states with $\mu \neq E_z$ are therefore given by
\begin{align}
    \ket{\mu<E_z,\Bz}&=\prod_{n=1}^{N_{max}(\mu)} \prod_{k_x=-\infty}^{\infty}d_{n,k_x}^\dagger \ket{\mathrm{vac}} \, , \\
     \ket{\mu>E_z,\Bz}&=\prod_{n=0}^{N_{max}(\mu)} \prod_{k_x=-\infty}^{\infty}b_{n,k_x}^\dagger \ket{\mathrm{vac}} \, ,
\end{align}
where one should pay special attention to the role of the $n=0$ \ac{LL}. Again, we insert these equations into Eq.~\eqref{eq:sigmaIIDef} and arrive for $\mu<E_z$ at
\begin{align}
    \sigma_{xy}^{II}=-\frac{e^2}{h} \sum_{n=1}^\infty \theta\lb -\mu+E_{n}^- \rb  \, ,
\end{align}
and for $\mu>E_z$ at
\begin{align}
    \sigma_{xy}^{II}=\frac{e^2}{h} \left[\theta\lb\mu-E_{n=0}\rb +\sum_{n=1}^\infty \theta\lb \mu-E_{n}^+ \rb \right] \, ,
\end{align}
where $E_{n=0}=-M+(B-D)/l_{\Bz}^2$. These steps must be repeated for all possible signs of $M,B$ and $D$, as well as one has to consider the additional cases for which $\abs{\Bz} > B_{\perp,triv}$. After a lengthy but straightforward calculation, a general expression for $\sigma_{xy}^{II}$ can be finally determined:
\begin{widetext}
\begin{align} \label{eq:sigmaII}
    \sigma_{xy}^{II} =
    -\frac{e^2}{2 h}   \left[\sgn{M-\frac{B}{l_{\Bz}^2}}+ \sgn{e \Bz \bar{\mu}} \right] \theta\lb\abs{\bar{\mu}}-\abs{M-\frac{B}{l_{\Bz}^2}}\rb -\sgn{e \Bz}\frac{e^2}{h} \sum_{n=1}^\infty \left[\theta\lb \mu-E^+_n\rb-\theta\lb -\mu +E^-_n\rb \right] \, ,
\end{align}
\end{widetext}
where $\bar{\mu}\equiv\mu+D/l_{\Bz}^2$.

Let us summarize the physical implications which we can derive from Eqs.~\eqref{eq:sigmaI} and~\eqref{eq:sigmaII}. The first term, $\sigma_{xy}^I$, is connected solely to the spectral asymmetry and is as such a signature of the parity anomaly. It is only nonzero if the system is for $\Bz=0$  a QAH insulator, i.e., $M/B>0$. We probe exclusively $\sigma_{xy}^I$, the `QAH regime', if the chemical potential is placed within
\begin{align} \label{eq:chernInsDiracGap}
    \abs{\mu+D/l_{\Bz}^2} \leq \abs{M-B/l_{\Bz}^2} \, .
\end{align}
We refer to this regime as the \textit{Dirac mass gap} because of its relation to the intrinsic bulk band gap of a Chern insulator. Given that $\mu$ is placed within the Dirac mass gap, the Hall conductivity is an even function of the magnetic field, i.e., $\sigma_{xy}^I (-\Bz)=\sigma_{xy}^I (\Bz)$. Since, in contrast, in the conventional QH phase, the Onsager relation implies that $\sigma_{xy}$ must be an odd function of $\Bz$~\citep{AshcroftBook}, we will refer to this characteristic signature of a QAH phase as `violation of the Onsager relation'  (for further discussions, see App.~\ref{app:Onsager}).  Moreover, Eqs.~\eqref{eq:sigmaI} and~\eqref{eq:chernInsDiracGap} highlight  a competition between the \textit{bare}  Dirac mass $M$ and the non-relativistic mass $B$. The $B$-parameter causes a decrease of the Dirac mass gap until it is eventually closed at $\Bz=B_{\perp,triv}$ above which $\sigma_{xy}^I=0$. In comparison, the $D$-parameter comes at the same level as the chemical potential and, hence, shifts the center of the Dirac mass gap in magnetic fields. The difference between the parameters arises because only $M$ and $B$ break parity symmetry at $\Bz=0$ (App.~\ref{app:symmetries}).

The second term in Eq.~\eqref{eq:totalHall}, $\sigma_{xy}^{II}$, contributes additionally to the total Hall conductivity only if the chemical potential is placed outside of the Dirac mass gap, so that extra \acp{LL} are filled/emptied with respect to the ground state. In contrast to Eq.~\eqref{eq:sigmaI}, each of these contributions  is related to a single \ac{LL}. Their origin is reflected by their $\sgn{e \Bz}$-dependence, seen in Eq.~\eqref{eq:sigmaII}. More precisely, for $\sgn{e \Bz}>0$ every conduction band \ac{LL} contributes $-e^2/h$ and every valence band \ac{LL} $+e^2/h$. The signs come in reverse when we flip the 
  direction of the magnetic field. Hence, Eq.~\eqref{eq:sigmaII} describes conventional QH physics, generated by the external magnetic field.

Finally, let us briefly comment on the connection between our findings and \ac{QED} in $d=2+1$ dimensions. A Chern insulator is equivalent to a (2+1)D Dirac equation for $B=D=0$ and $A=1$ (limit has to be taken before regularization). In this case, only the unpaired $n=0$ LL contributes to the spectral asymmetry and $\etaH = n_0 \, \sgn{e \Bz} \sgn{M}$ \cite{Niemi83,Boyanovsky86}. This leaves us with a half-quantized Hall response and, therefore, with a fractional ground state charge \cite{Haldane88,Schakel91}. In contrast, a \ac{QAH} insulator, as described by Eq.~\eqref{eq:sigmaI} exhibits an integer quantized Hall response and, hence, an integer-valued ground state charge. The momentum-dependent mass term $B$ contributes additionally to the spectral asymmetry and acts as if there is an effective partner of the otherwise unpaired $n=0$ \ac{LL}. The field theoretical aspect of this point is discussed in detail in Ref.~\cite{Tutschku20}.

%%%%%%%%%%%%%%%%%%%%%%%%%%%%%%%%%%%%%%%%%%%%%%%%%%
%%     MODEL - BULK BOUNDARY CORRESPONDENCE 
%%%%%%%%%%%%%%%%%%%%%%%%%%%%%%%%%%%%%%%%%%%%%%%%%%
\section{Hall conductivity: Edge perspective} \label{sec:bulkboundary}

%%%%%%%%%%%%%%%%%%%%%%%%%%%%%%%%%%%%%
\begin{figure*}[t]
    \includegraphics[width=.9\textwidth]{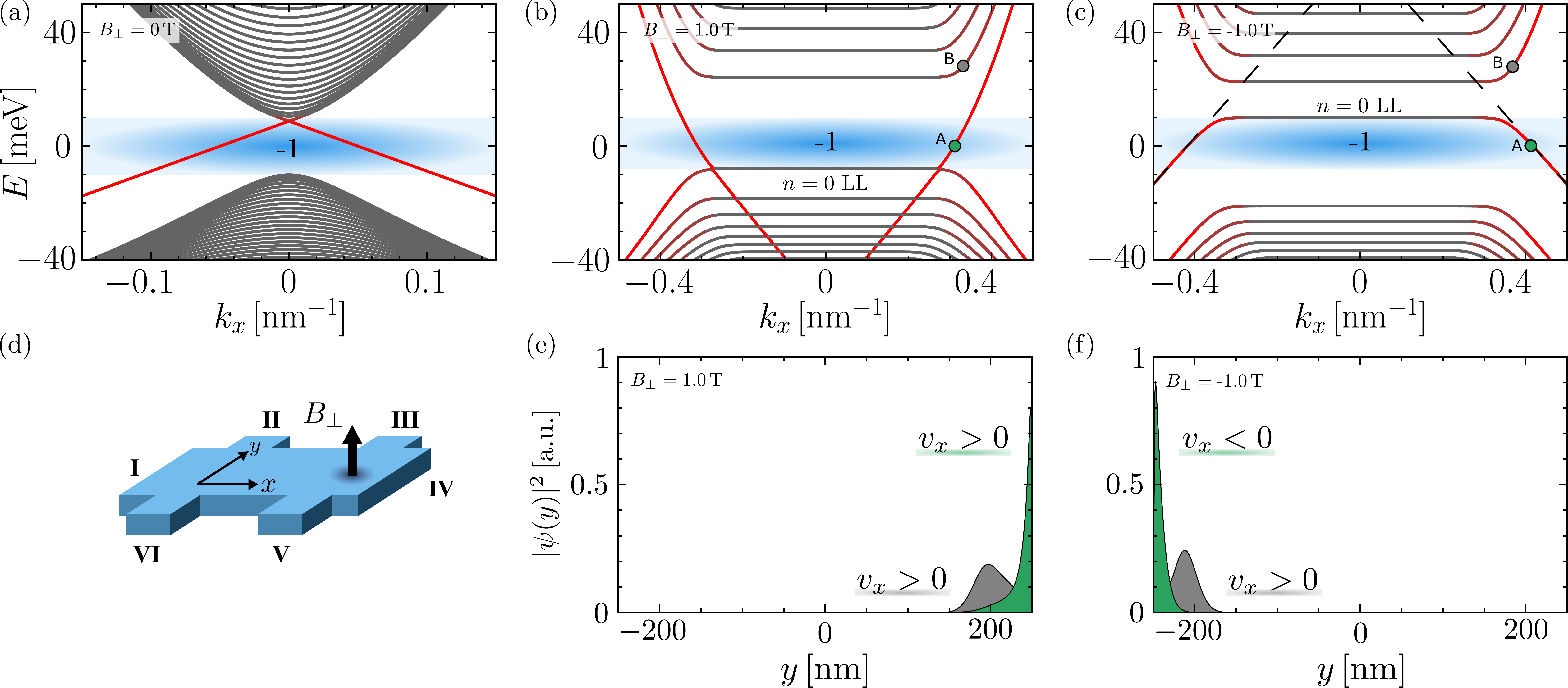}
    \caption{\label{fig:llSymmetryMag}
    Landau level spectrum of Chern insulator with $M=-10 \meV$, $B=-685\meV \nm^2$, $D=-600\meV \nm^2$, and $A=365 \meV \nm$ mapped on a lattice with $L_y = 500 \nm$, and $a=1 \nm$ for a magnetic field of (a) $0 \tesla$, (b) $1 \tesla$, and (c) $ -1 \tesla$. Color code displays the wave function localization. Edge (bulk) states are depicted in red (gray).  The blue shaded area marks the  Dirac mass gap [Eq.~\eqref{eq:chernInsDiracGap}] characterized by  $\sigma_{xy}(-\Bz) = \sigma_{xy} (\Bz)$, where Chern numbers are explicitly shown. In (c), dashed line indicates evolution of QAH edge states in the conduction band before hybridizing with bulk states.  (d) Sketch of conventional six-terminal Hall bar. In (e) and (f), probability density of two wave functions is depicted corresponding to point A (green dot) and B (gray dot) in the spectrum shown in (b) and (c), respectively. Sign of Fermi velocity is highlighted. }
\end{figure*}
%%%%%%%%%%%%%%%%%%%%%%%%%%%%%%%%%%%%%%

The bulk-boundary correspondence implies that topological edge states must exist at the boundary between  topologically nontrivial and trivial regimes \cite{Callan85,Hasan10}. The number of edge channels is determined by the absolute value of the filling factor $\abs{\nu}=|\sigma_{xy}|h/e^2$ and their chirality is given by the sign of each individual contribution $\nu_i=\pm1$, with $\nu=\sum_i \nu_i$. For a QAH insulator in magnetic fields, there are two sources giving rise to topological edge channels: The intrinsic QAH topology and the external magnetic field. We now map a QAH insulator  [Eq.~\eqref{eq:chernInsulator}] on a lattice \cite{Scharf12} and study the properties of the associated, distinct topological edge states. Analogously to the previous bulk calculations, we keep periodic boundary conditions in the $x$-direction, but require the wave functions to vanish at $y=\pm L_y/2$.  The procedure of mapping the Hamiltonian on the lattice introduces a hard cut-off in momentum space with $k_{max} \sim a^{-1}$, where $a$ is the lattice constant, i.e., the lattice acts as the necessary regulator \cite{Kogut83}. In comparison to the continuum theory, this guarantees a finite number of degrees of freedom, so that no further regularization, like employing a heat-kernel, is needed to determine physical observables like the charge or the Hall conductivity \footnote{Physical observables should not depend on the chosen regularization scheme.}. 

Let us now focus on a QAH insulator with $M,B<0$. In the absence of a magnetic field, the Dirac point is located in the bulk gap at $E_z=-M D / B$ \cite{Zhou08}, where the system is at half-filling. The charge neutrality point coincides therefore with the Dirac point, employed already in Sec.~\ref{sec:brokenPHS}. The corresponding band structure at $\Bz=0$ is shown in Fig.~\ref{fig:llSymmetryMag}(a), for which the color code reflects the spatial wave function localization in the $y$-direction. Bulk LLs are depicted in gray, while edge states are displayed in red. The chosen parameters are similar to those of (Hg,Mn)Te quantum wells \cite{Koenig08,Beugeling12}.

Gradually increasing now the magnetic field pushes the QAH edge states and, associated, the Dirac point into the valence band for $\sgn{e \Bz}>0$, or into the conduction band  for $\sgn{e \Bz}<0$ \cite{Zhou08,Boettcher19}.
To make this more explicit, two Chern insulator spectra are shown for out-of-plane magnetic fields of $1 \tesla$ and $-1 \tesla$ in Figs.~\ref{fig:llSymmetryMag}(b) and~(c), respectively.  It is apparent that for the given system parameters and independent of the magnetic field direction, edge states traverse the Dirac mass gap [cf. Eq.~\eqref{eq:chernInsDiracGap}], which is marked by the blue area in Figs.~\ref{fig:llSymmetryMag}(a)--(c). Recall that the Dirac mass gap signifies the regime where the Hall conductivity is an even function of the magnetic field. Based on the principle of bulk-boundary correspondence, we can infer that these states must be uniquely related to $\sigma_{xy}^I$, Eq.~\eqref{eq:sigmaI}, since this is the only term which can contribute within the Dirac mass gap. Ergo, these states are the descendants of the QAH edge states and exist only if $M/B>0$ and $\Bz < B_{\perp,triv}$.

The QAH edge states can also continue to exist outside of the Dirac mass gap, since they are not bound by a Heaviside step function  [cf. Eq.~\eqref{eq:sigmaI}]. This is clearly observed for positive magnetic fields as shown in Fig.~\ref{fig:llSymmetryMag}(b), where counterpropagating (helical-like) QH and QAH edge states coexist in the valence band. To be more precise, for $k_x>0$, all QH edge states exhibit a negative Fermi velocity in the valence band, while the QAH edge state has a positive Fermi velocity.  However, there is no clear signature of QAH edge states outside of the Dirac mass gap  for negative magnetic fields, as shown in Fig.~\ref{fig:llSymmetryMag}(c). In the latter case, this is because the QAH edge states are strongly hybridized  with conduction band states. To illustrate this, we indicated the QAH edge states  before hybridizing with QH states by dashed lines in Fig.~\ref{fig:llSymmetryMag}(c). We explain the parameter dependence of this process and its physical implications in the next section.

Let us now focus at first on the signatures of the QAH edge states within Dirac mass gap. In this regime, QH edge states cannot exist [see Eqs.~\eqref{eq:sigmaI} and~\eqref{eq:sigmaII}] so that the QAH edge states are protected from hybridization.
In particular, we focus on how the violation of the Onsager relation is connected to the QAH edge states. 
To this end, the wave functions for two selected points are shown in Figs.~\ref{fig:llSymmetryMag}(e) and~(f), corresponding to the two marked points in  Figs.~\ref{fig:llSymmetryMag}(b) and~(c), for the two magnetic field configurations. \textit{Point A} marks a QAH edge state inside of the Dirac mas gap, while \textit{point B} marks a conventional QH edge state outside of the Dirac mass gap. For $\Bz>0$, edge states with positive momenta  are localized at the top edge ($y=L_y/2$) of our stripe geometry [Fig.~\ref{fig:llSymmetryMag}(e)], while they localize at the bottom edge ($y=-L_y/2$) for $\Bz<0$ [Fig.~\ref{fig:llSymmetryMag}(f)]. Flipping the magnetic field direction results therefore in changing the spatial localization of edge states at given $k_x$.

The associated Hall resistance  $R_H$ can be now computed employing the Landauer-B\"uttiker formalism.  For a six terminal set-up, schematically depicted in Fig.~\ref{fig:llSymmetryMag}(d), it follows that \cite{Chen12,Boettcher19}:
\begin{align} \label{eq:LBuettiker}
    R_H=\frac{h}{e^2} \frac{T_c-T_a}{T_c^2-T_a T_c+T_a^2} \, ,
\end{align}
where $T_c=T_{i+1 \leftarrow i}$ is the transmission probability in clockwise direction, i.e., from the $i$-th to the $(i+1)$-th contact; $T_a=T_{i\leftarrow i+1}$ is the transmission probability in anticlockwise direction. We consider first the QAH case in which the chemical potential is placed at the \textit{point A} in Fig.~\ref{fig:llSymmetryMag}(b). Since the Fermi velocity $v_x=\hbar^{-1} \partial E / \partial k_x$ is positive for $k_x>0$ and the wave function is located at the top edge of our stripe geometry, there is only one chiral edge channel propagating clockwise along the edges of the Hall bar. This amounts to $T_c=1$ and $T_a=0$ resulting in $R_H = h/e^2$. 

In comparison, placing the chemical potential at the \textit{point A} in Fig.~\ref{fig:llSymmetryMag}(c), the wave function of the QAH edge state  for $k_x>0$ is located at the bottom edge and exhibits a negative Fermi velocity, i.e., both, the edge localization and the Fermi velocity flip sign for $\Bz \rightarrow -\Bz$. The two effects combined yield again a clockwise propagating edge state which exhibits the same transmission probabilities and, hence, the same Hall resistance $R_H = h/e^2$, as the QAH edge state in Fig.~\ref{fig:llSymmetryMag}(b). This originates from the fact that, in both cases, the same QAH edge state is probed whose chirality is defined by the intrinsic Chern number and not by the magnetic field. As a result, the Hall conductivity in the Dirac mass gap is an even function of the magnetic field and, therefore, violates the Onsager relation [cf. App.~C]. This property holds as long as QAH edge states are allowed to bridge this gap, i.e., for $\Bz<B_{\perp,triv}$ and $M/B>0$.

This property is clearly different from the QH edge states induced by the external magnetic field outside of the Dirac mass gap. While edge states still change their spatial localization, their Fermi velocity remains the same for $\Bz \rightarrow -\Bz$. As a result, the transmission probability of each QH edge state in the conduction band changes from $(T_c,T_a)=(1,0)$ to $(0,1)$ for $\Bz \rightarrow -\Bz$ and their contribution to the total Hall resistance [Eq.~\eqref{eq:LBuettiker}] changes from $+h/e^2$ to $-h/e^2$. The signs are opposite for valence band \acp{LL}, where the Hall resistance changes from $-h/e^2$ to $+h/e^2$.

%%%%%%%%%%%%%%%%%%%%%%%%%%%%%%%%%%%%%%%%%%%%%%%%%%
%%     MODEL - ROLE OF D PARAMETER 
%%%%%%%%%%%%%%%%%%%%%%%%%%%%%%%%%%%%%%%%%%%%%%%%%%
\section{Influence of broken PH symmetry on QAH edge states in magnetic fields} \label{sec:hybDiscussion}
We now turn the focus to properties of the QAH edge states outside of the Dirac mass gap. In particular, we clarify why, in this regime for the given set of parameters, QAH edge states are only clearly visible for $\Bz>0$ [Fig.~\ref{fig:llSymmetryMag}(b)] and not for $\Bz<0$ [Fig.~\ref{fig:llSymmetryMag}(c)]. Let us first start with the case of positive magnetic fields.

\begin{figure}[!b]
    \includegraphics[width=.98\columnwidth]{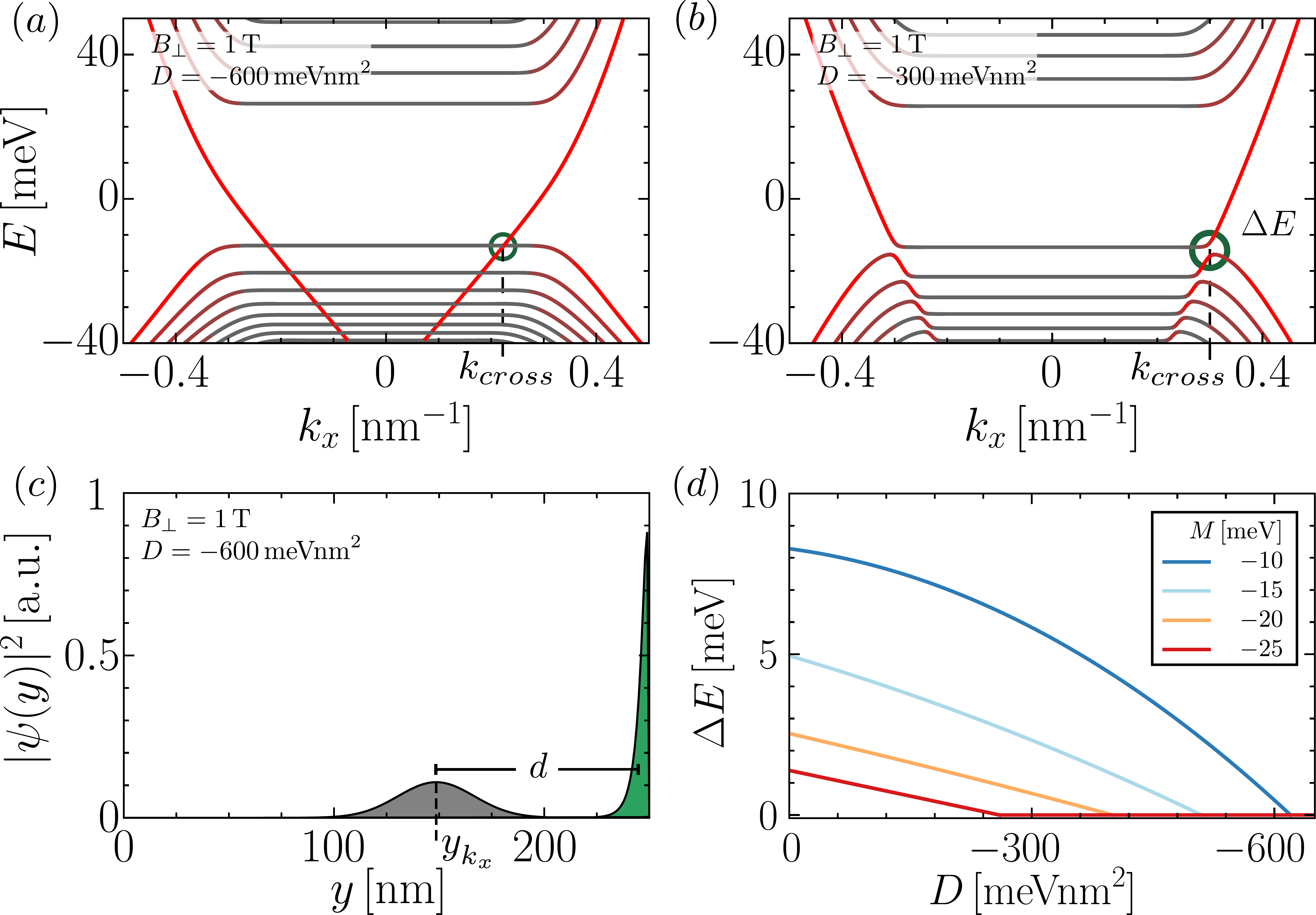}
    \caption{\label{fig:waveFunctionOverlap} Effect of $D$-parameter on band structure of Chern insulator for $M=-15 \meV$, $A=365\meVnm$, $B=-685 \meVnm^2$, (a) $D=-600 \meVnm^2$, and (b) $-300\meVnm^2$ at $\Bz=1 \tesla$. (Anti)crossing at $k_{cross}$  between \ac{QAH} edge state and $n=0$ \ac{LL} is marked by green circle. (c) Probability density of wave functions that are associated to $k_{cross}$ in (a) between \ac{QAH} edge state (green) and Gaussian wave function of $n=0$ \ac{LL} (gray).  The Gaussian probability density is centered at $y \approx 145 \, \mathrm{nm}$. The distance between the centers is given by $d$. (d) Hybridization gap between  QAH edge state and $n=0$ \ac{LL} as function of $D$-parameter for various Dirac masses $M$. The numerical resolution determining the gap is $\Delta E = \pm 0.1 \meV$.  }
\end{figure}

As we have stressed earlier,  QAH states can hybridize with QH  states outside of the Dirac mass gap, since they are not protected by symmetry. The crossing (up to finite size gaps) between the QAH edge states and the $n=0$ \ac{LL}, as it is observed in Fig.~\ref{fig:waveFunctionOverlap}(a), at  $k_{cross}$ is instead protected by differently localized wave functions [Fig.~\ref{fig:waveFunctionOverlap}(c)]. This situation is however not generic as a comparison with a Chern insulator with smaller $D$-parameter shows [see avoided crossings in Fig.~\ref{fig:waveFunctionOverlap}(b)].
%%%%%%%%%%%%%%%%%%%%%%%%%%%%%%%%%%%%%
%%%%%%%%%%%%%%%%%%%%%%%%%%%%%%%%%%%
To analyze this process quantitatively, we plot in Fig.~\ref{fig:waveFunctionOverlap}(d) at $\Bz=1\tesla$ the hybridization gap $\Delta E$ of the QAH edge states and the $n=0$ \ac{LL}  as a function of $D$ for various Dirac masses $M$. The results show that increasing the absolute value of $M$ and $D$ increases the regime in which $\Delta E$ drops to zero (numerical resolution: $\Delta E = \pm 0.1 \meV$). This means that a strong PH asymmetry protects the edge states from hybridization with bulk LLs. In the following, we label the critical magnetic field  above which $\Delta E\neq 0$, i.e., a hybridization gap starts to form, by $B_{\perp,hyb}$.

 Physically, the behavior of  $B_{\perp,hyb}$ in Fig.~\ref{fig:waveFunctionOverlap}(d) can be intuitively understood noting that this critical field is basically determined by the parameter dependence of $k_{cross}$. This is because $k_{cross}$ is intimately linked to the wave function overlap of QAH edge states and QH states. We can find an analytic estimate for $k_{cross}$ based on the energetic position of the $n=0$ \ac{LL}, Eq.~\eqref{eq:LLenergiesD} and the dispersion of the QAH edge states in magnetic fields. The latter is given by \cite{Zhou08}
 \begin{align}
    E_{edge,s}^{\pm} (k_x)=E_z -s \mu_B g_{eff}^{} \Bz \pm \hbar v_x k_x \, ,
\end{align}
where $g_{eff}^{} \approx m_0 v_x L_y / \hbar$ \footnote{The approximation is valid for $L_y\gg\lambda_{1,2}$, where $\lambda_{1,2}$ are the decay length scales of the edge state.}, $v_x=A\sqrt{(B^2-D^2)/B^2}/\hbar$, and $s=\pm$ labels the edge states of the spin up ($+$) or down ($-$) Chern insulator, respectively. In the present case (spin up), the two crossing points are determined by solving $E_{n=0}(\Bz)=E_{edge,\spinup}^+ (k_{cross})$ for $k_{cross}$.

Hybridization is almost absent if $\abs{k_{cross}}\ll k_{max}$ ($\Delta E \approx 0$), where $2k_{max}= L_y \abs{ e\Bz} /\hbar$ is the maximal width of a bulk \ac{LL}. This is due to the fact that the QAH edge states are exponentially localized at the edges, while the  Gaussian wave functions of the QH states are each centered at $y_{k_x}^{}=l_{\Bz}^2 k_x$ and have a standard deviation of $\sigma= l_{\Bz}^{}$.
The probability densities of these wave functions are shown exemplary in Fig.~\ref{fig:waveFunctionOverlap}(c) for the marked crossing point in Fig.~\ref{fig:waveFunctionOverlap}(a). In this case, coexisting, counterpropagating QH and QAH edge states are observed in the valence band.

\begin{figure}[!t]
    \centering
    \includegraphics[width=.98\columnwidth]{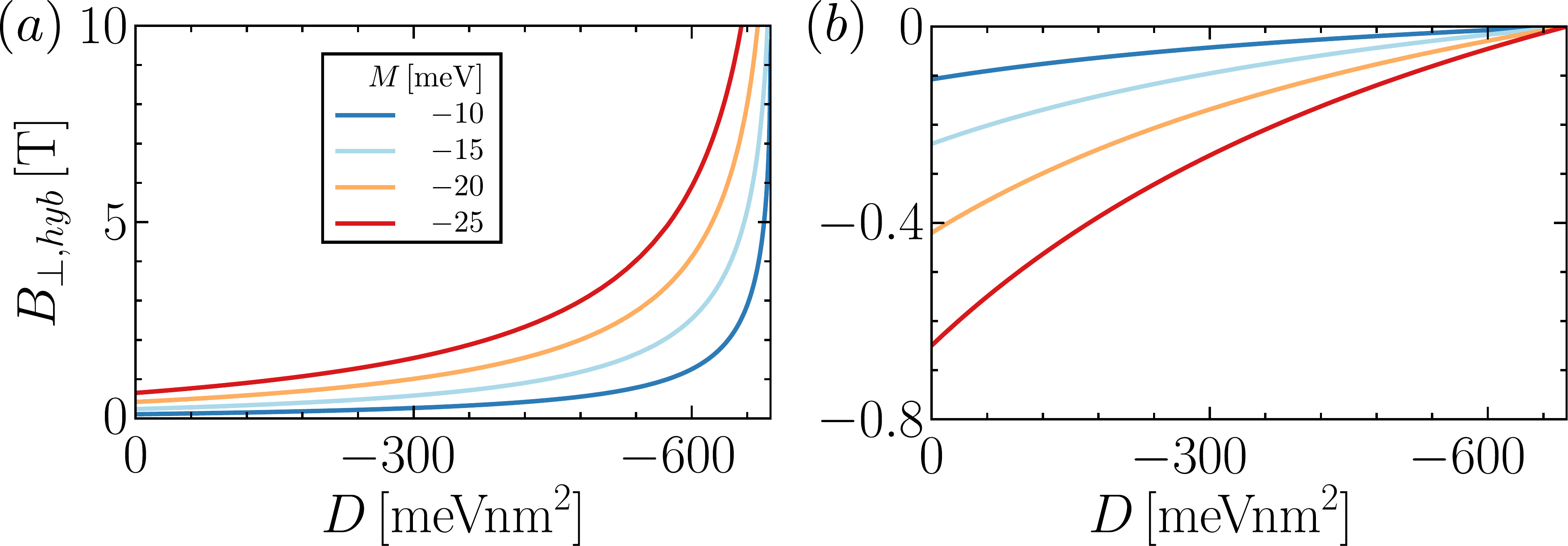}
    \caption{\label{fig:BHybEvol}
   Evolution of $B_{\perp,hyb}$ as function of $D$-parameter for (a) positive and (b) negative magnetic fields with $M=-10,-15,-20$, and $-25 \meV$ [see plot legend in (a)]. $B_{\perp,hyb}$ increases with increasing absolute value of $M$ and $D$. The fitting parameter $c \approx 1.06$ (see text for further discussion). }
\end{figure}

A hybridization gap $\Delta E$ starts to form only if the distance $d$ between the wave function centers gets of the order of the standard deviation $\sigma$, i.e., if $d<d_{crit}=c \sigma$. Here, $c>0$ is a fitting parameter which can be adjusted to gain good agreement with the numerical results (typically, $c\sim1$).  Assuming that the center of the QAH edge state lies approximately at the sample edge, the distance between the wave function centers is roughly given by $d=L/2-y_{k_x}^{}$.  Based on these rough estimates, we find that hybridization becomes relevant for
\begin{align}
    k_{cross}\gtrsim(L/2-c\sigma)l_{\Bz}^{-2}\equiv k_{hyb} \, .
\end{align}
An expression for the corresponding critical magnetic field is found by solving $E_{n=0}(B_{\perp,hyb})=E_{edge,\spinup}^+ (k_{hyb})$ for $B_{\perp,hyb}$, but we refrain here from showing the full analytic expression as it is very long and cumbersome. To gain some insight in the parameter dependence of $B_{\perp,hyb}$,  we Taylor expand the analytical expression up to first order in $D$, resulting in
\begin{align} \label{eq:bhyb}
    B_{\perp,hyb}\approx B_0\lb \sgn{ e \Bz } + \frac{2 c A}{B\sqrt{4MB +c^2A^2}}D\rb \, ,
\end{align}
where
\begin{align}
    B_0=\frac{\hbar}{e B^2}\left[ M B + \frac{1}{2} c A\lb c A-\sqrt{4 M B+c^2 A^2}\rb \right] \, .
\end{align}
The full analytic behavior of $B_{\perp,hyb}(D)$ for positive and negative magnetic fields is shown in Fig.~\ref{fig:BHybEvol}(a) and (b), respectively.

%%%%%%%%%%%%%%%%%%%%%%%%%%%%%%%%%%%%%%%%%%
%%%%%%%%%%%%%%%%%%%%%%%%%%%%%%%%%%%%%%%%%%%%
 %%%%%%%%%%%%%%%%%%%%%%%%%%%%%%%%%%%%%%%%%%
\begin{figure}[!b]
    \centering
    \includegraphics[width=.98\columnwidth]{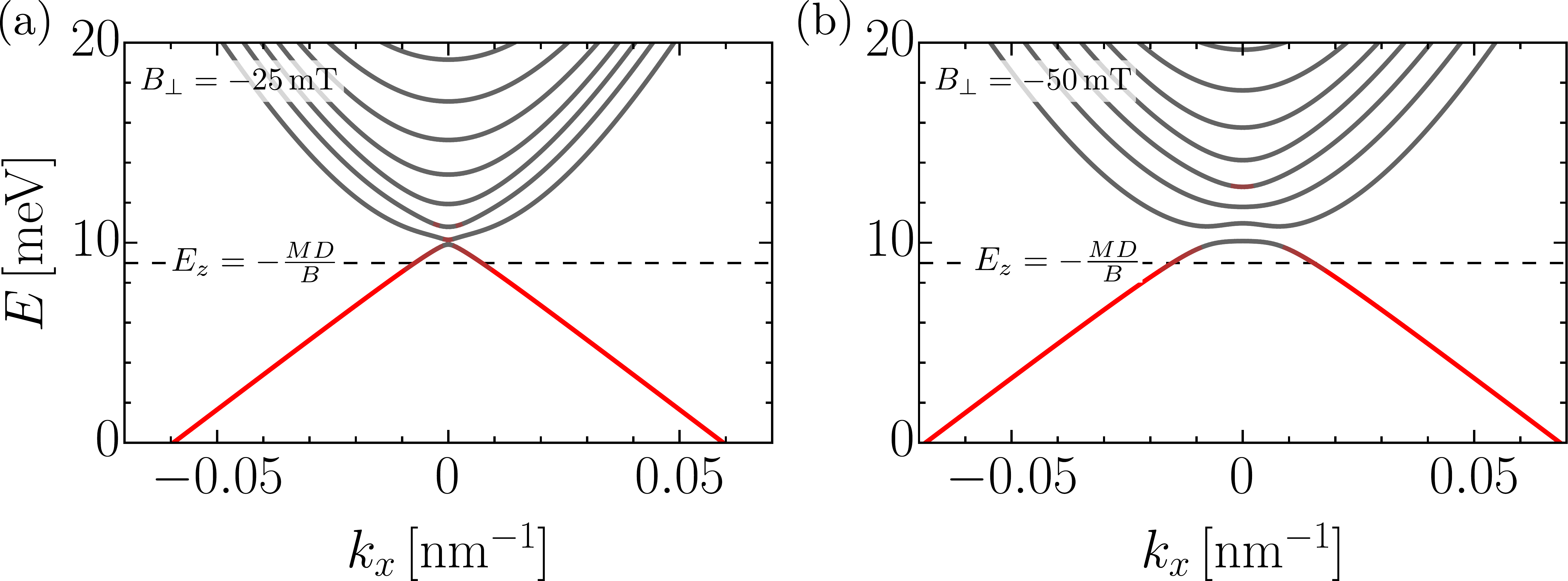}
    \caption{\label{fig:llNegMag} Band structures of Chern insulator for small, negative magnetic fields of (a) $-25 \mTesla$ and (b) $-50 \mTesla$ are shown. QAH edge states are pushed into the conduction band, where they strongly hybridize with bulk bands. Dashed line marks energy of Dirac point for $\Bz=0$. For a better visibility, the energy range is limited from  $E=0$ to $20 \meV$.
    }
\end{figure}
%%%%%%%%%%%%%%%%%%%%%%%%%%%%%%%%%%%%%%%%%%%%

Taken together, Eq.~\eqref{eq:bhyb} and Fig.~\ref{fig:BHybEvol} show that, for $\Bz>0$, the absolute value of $B_{\perp,hyb}$ increases when $D$ approaches $B$, i.e., a strong PH asymmetry protects the edge channels from hybridization. This is in contrast to the case of negative magnetic fields, where a strong PH asymmetry causes already at a few mT a strong hybridization of QH and QAH edge states.
This is in accordance with the band structure calculations presented in Fig.~\ref{fig:llNegMag}, which show that already magnetic fields of $\Bz<-25\mTesla$ are sufficient to push the Dirac point into the conduction band and to cause large hybridization gaps.
We can therefore attribute the  difference in the appearance of the QAH edge states in Fig.~\ref{fig:llSymmetryMag}(b) and~(c) at $\Bz=\pm 1\tesla$ to a strong PH asymmetry.  Let us finally reiterate that this does not affect any signature of the QAH edge states within the Dirac mass gap, where they remain to be protected from hybridization.

Note that in real 2D topological insulators, which are described at low energies by the \ac{BHZ} model, like (Hg,Mn)Te \cite{Bernevig06} or InAs/GaSb bilayers \cite{LiuInAs08}, numerical deviations from these results might occur due to the natural limitations of the model. In particular, deviations can arise from employing the low-energy \ac{BHZ} model and assuming an impurity-free system. Both assumptions can affect the explicit form of the wave functions, the position of the Dirac point \cite{Wimmer18,Saquib19} and, hence, alter the hybridization of QH and QAH edge states. Recently, it has been also argued that the use of hard wall boundary conditions can affect the position of Dirac points \cite{Klipstein18,Gioia19}.
Independent of these possible contributions, Eq.~\eqref{eq:sigmaI} shows that the critical magnetic field, above which a hybridization gap must start to form, is given by $B_{\perp, triv}$ [Eq.~\eqref{eq:trivField}]. This is due to the fact that edge states are then no longer allowed to traverse the Dirac mass gap. In Ref.~\cite{Boettcher19}, we proposed possible routes towards verifying their existence in transport experiments. This includes a special type of charge pumping in increasing magnetic fields and a characteristic transport behavior which is associated to the counterpropagating QAH and QH edge states in the presence of charge puddles. 

%%%%%%%%%%%%%%%%%
\begin{figure}[!t]
    \centering
    \includegraphics[width=0.99\columnwidth]{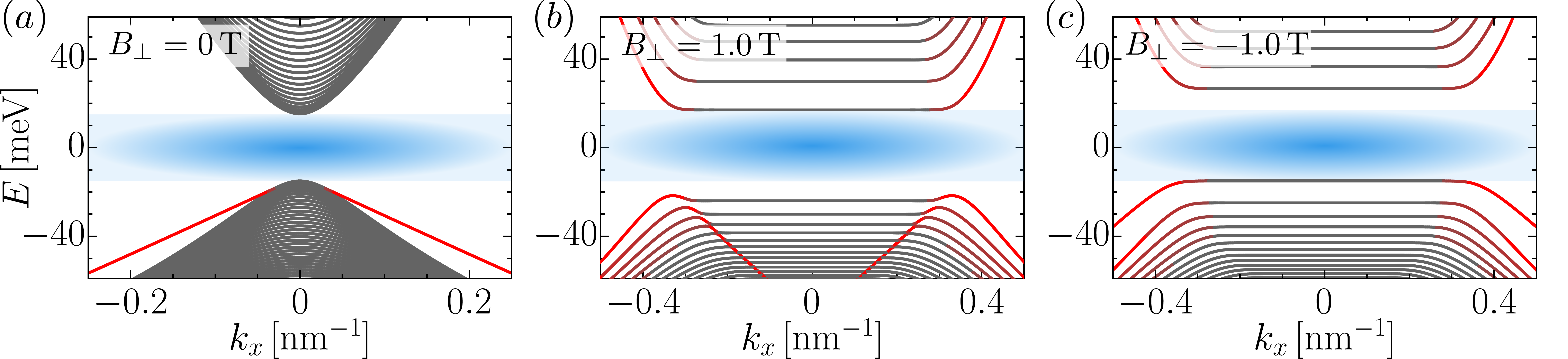}
    \caption{\label{fig:trivband} Band structure of trivial Chern insulator for (a) $\Bz=0\tesla$, (b) $1\tesla$, and (c) $-1\tesla$. Same parameters as in Fig.~\ref{fig:llSymmetryMag} are used except for $M=+15\meV$. A trivial edge state is observed in (a) which does not traverse the Dirac mass gap. By virtue of the trivial topology, there are no edge states in the Dirac mass gap (blue regime) in magnetic fields. 
}
\end{figure}
%%%%%%%%%%%%%%%%

We conclude this section by comparing our findings with a topologically trivial Chern insulator ($M/B<0$) with $D\neq0$.  In this case, the Hall conductivity $\sigma_{xy}$ is solely determined by $\sigma_{xy}^{II}$, since the two contributions to the spectral asymmetry cancel each other, resulting in $\sigma_{xy}^I=0$. Even though the system is of trivial topology, edge states, depicted in red, can exist outside of the Dirac mass gap for $\Bz=0$, as shown in Fig.~\ref{fig:trivband}(a). This paradoxical extension of edge states to the topologically trivial regime can be explained by an emergent, approximate chiral symmetry, provided that \ac{PH} symmetry is violated at $\Bz=0$ \cite{Candido18}. But in stark contrast to  QAH edge states, trivial edge states can never enter into the Dirac mass gap, even for $\Bz \neq 0$ [cf. Fig.~\ref{fig:llSymmetryMag} and~\ref{fig:trivband}].

%%%%%%%%%%%%%%%%%%%%%%%%%%%%%%%%%%%%%%%%%%%%%%%%%%
%%     MODEL - BHZ MODEL
%%%%%%%%%%%%%%%%%%%%%%%%%%%%%%%%%%%%%%%%%%%%%%%%%%
\section{BHZ model}\label{sec:BHZ}

%%%%%%%%%%%%%%%%
\begin{figure}[!b]
    \centering
    \includegraphics[width=.98\columnwidth]{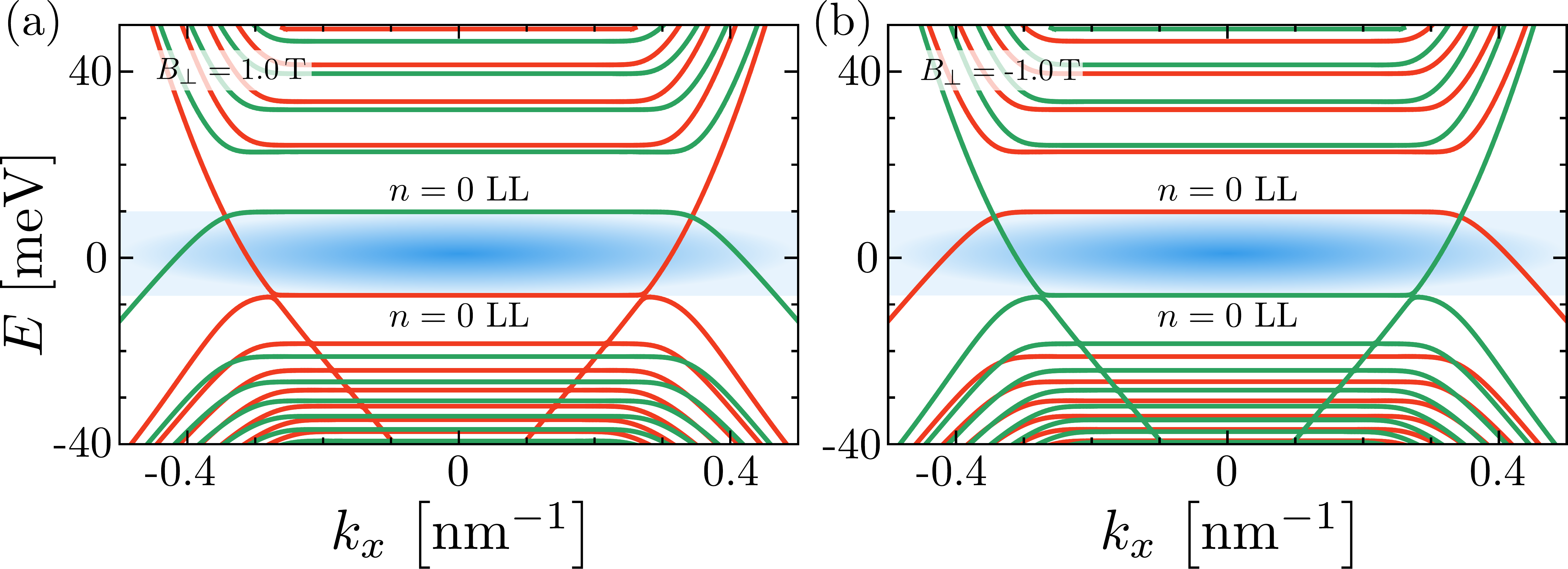}
    \caption{\label{fig:llBHZSymmetryMagQSH}Landau level spectrum of \ac{BHZ} model without Zeeman or exchange interaction terms for (a) $\Bz=1\tesla$ and (b) $-1\tesla$. The color code corresponds to the spin up (orange) and the spin down (green) block. Model parameters are the same as in Fig.~\ref{fig:llSymmetryMag}. The Dirac mass gap (blue shaded area) is characterized by counterpropagating edge states. The spectrum is invariant under $\Bz\rightarrow -\Bz$ with the exception that the spin up and down block interchange their role.}
\end{figure}
%%%%%%%%%%%%%%%%%%%%

We now go back to a \ac{QSH} insulator described by the full \ac{BHZ} model. To this end, we re-introduce the spin index to distinguish the two  spin blocks [Eq.~\eqref{eq:fullBHZ}]. 
Following our discussion below Eq.~\eqref{eq:unitaryTrafoBHZ}, we see that  $\etaSpin{\spindown}=-\etaSpin{\spinup}$, where $\etaSpin{\spinup}\equiv\etaH$ which is given by Eq.~\eqref{eq:spectralAsymPHS}. The spectral asymmetry is therefore determined by
\begin{align} \label{eq:qshEta}
\eta_{BHZ}^{}(\Bz)=\etaSpin{\spinup}+\etaSpin{\spindown}=0 \, .
\end{align}
This  is accompanied by a vanishing Hall conductivity $\sigma_{xy}$ in the Dirac mass gap:
\begin{align} \label{eq:qshCond}
\sigma_{xy}=\sigma_{xy,\spinup}^I+\sigma_{xy,\spindown}^I 
\end{align}
where, according to Eq.~\eqref{eq:sigmaI},
\begin{align} \label{eq:qshCondPart}
\sigma_{xy,s}^I= \pm \frac{e^2}{2h}\!\left[\sgn{M-\frac{B}{l_{\Bz}^2}}\!+\sgn{B}\right] \, . 
\end{align}
Here, $s={\spinup,\spindown}$ corresponds to $\pm$, respectively.
That in this case $\sigma_{xy}=0$ is a consequence of the parity and \ac{TR} symmetry of the full model at $\Bz=0$. However, we can consider the odd combination of both spin blocks \cite{Semenoff84} to distinguish this state from a trivial insulator:
\begin{align} \label{eq:etaSpinBHZ}
    \eta_{BHZ}^S(\Bz) = \etaSpin{\spinup}-\etaSpin{\spindown} \, ,
\end{align}
which is nonzero for $M/B>0$ and $\Bz<B_{\perp,triv}$. This concept is related to the nontrivial spin Hall conductivity of a QSH insulator \cite{Qi06}:
\begin{align}
    \sigma_{xy}^S=\sigma_{xy,\spinup}^I-\sigma_{xy,\spindown}^I \, ,
\end{align}
which takes on quantized values in the ballistic regime if $M/B>0$. This relation  highlights the fact that the nontrivial spin Hall conductivity of a QSH insulator can be also interpreted in the language of the parity anomaly \cite{Semenoff84}, since it is related to the topological quantity  $\eta_{BHZ}^S$. For $\Bz>B_{\perp,triv}$, Eq.~\eqref{eq:etaSpinBHZ} drops to zero  corresponding to the point at which the spectral asymmetry of each spin block vanishes. Notably, $B_{\perp,triv}$ coincides exactly with the critical field, at which the two spin polarized $n=0$ \acp{LL} cross \cite{Scharf12}. This demonstrates that the information about the band inversion is contained in each individual Chern insulator.

For completeness, we show in Fig.~\ref{fig:llBHZSymmetryMagQSH}(a) and (b) the spectrum of a QSH insulator, described by the BHZ model, for magnetic fields of $1\tesla$ and $-1\tesla$, respectively. The color code marks the two spin blocks. A pair (per edge) of counterpropagating edge states traverses the Dirac mass gap (blue area), which is a hallmark of the underlying QSH topology \cite{Koenig08,Tkachov10}. Comparing Fig.~\ref{fig:llBHZSymmetryMagQSH}(a) and (b), we see that the spectrum remains unaltered for $\Bz\rightarrow -\Bz$, with the exception that the spin up and down block interchange their role. This property of the spectrum can be understood noting that  reversing the  magnetic field direction is equivalent to a parity transformation, as both processes effectively flip the sign of $M$ and $B$ and, therefore, interchange the two spin blocks.
This result also explains the salient asymmetry between the appearance of the QAH edge states for the two spin directions. In Fig.~\ref{fig:llBHZSymmetryMagQSH}(a), the  strong PH asymmetry protects the QAH edge states of the spin-up block from hybridization, while it causes hybridization gaps for the spin-down block already at very small, positive magnetic fields. As observed in Fig.~\ref{fig:llBHZSymmetryMagQSH}(b), the situation is in reverse, when we flip the magnetic field direction [cf. to discussion in Sec.~\ref{sec:hybDiscussion}].

%%%%%%%%%%%%%%%%%%%%
 \begin{figure}[!b]
    \centering
    \includegraphics[width=0.98\columnwidth]{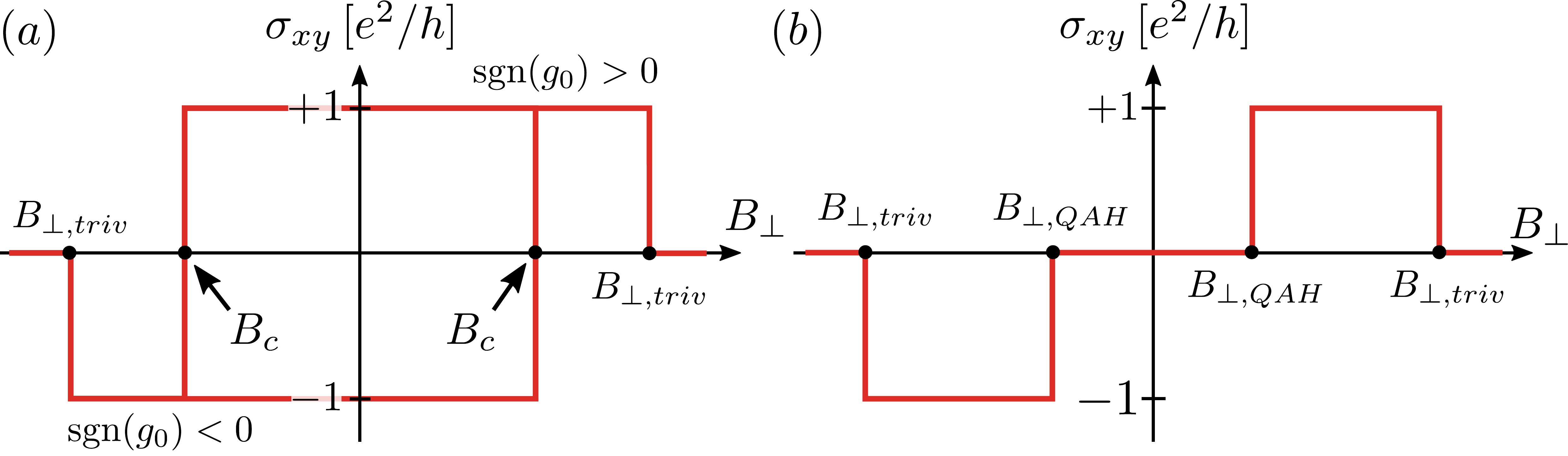}
    \caption{\label{fig:paraVsFerro}
    Sketch of $\sigma_{xy}$ is shown for (a) ferromagnetic, and (b) paramagnetic exchange interaction (Zeeman term) as function of external magnetic field $\Bz$ at a constant chemical potential (placed within the Dirac mass gap).   (a) Sign of $\sigma_{xy}$ is determined by  polarization direction of magnetic domains ($g_0$) of the ferromagnet which is here supposed to follow a hysteresis. Signs of $g_0$ are indicated. The Hall conductivity can switch its sign at the coercive field $B_c$.  (b) In the case of a paramagnet, a finite magnetic field $B_{\perp,QAH}$ is needed to overcome the Dirac mass gap of one of the two spin blocks, resulting in a nonzero $\sigma_{xy}^I$. We assumed for this paramagnetic case that $g_0>0$.  In both cases, the Hall conductivity vanishes at $\Bz > B_{\perp,triv}$, where $\eta_{BHZ}^{}=0$. }
\end{figure}
%%%%%%%%%%%%%

Let us now introduce additionally a Zeeman or an exchange term of the following form $H_{s} = \sigma_0 \otimes \tau_z \, g(\Bz)$, given in the basis of Eq.~\eqref{eq:unitaryTrafoBHZ}. 
This term can be easily incorporated  by replacing in our results $M \rightarrow M \pm g(\Bz)$ for the spin up ($+$)  and spin down block ($-$), respectively. The two contributions to the total Hall conductivity $\sigma_{xy}$, which are exclusively determined by the spectral asymmetry in the Dirac mass gap, are therefore given by [cf. Eq.~\eqref{eq:qshCondPart}]
\begin{align} \label{eq:qahCondPart}
\sigma_{xy,s}^I= \frac{\pm e^2}{2h}\!\left[\sgn{M \pm g(\Bz)-\frac{B}{l_{\Bz}^2}}\!+\sgn{B}\right]\! 
\end{align}
where $s={\spinup,\spindown}$ corresponds to $\pm$, respectively. The particular importance of exchange and Zeeman terms originates from the fact, that they can drive a \ac{2D} topological insulator from the \ac{QSH} into the \ac{QAH} phase \cite{Liu08}. In the following, we are going to discuss two cases: Firstly, $H_s$ describes a ferromagnetic exchange interaction, and, secondly, a paramagnetic exchange (equivalently, Zeeman) interaction  \footnote{In a realistic material, a Zeeman term will always come on top of the either paramagnetic or ferromagnetic exchange interaction}.

%%%%%%%%%%%%%%%%%%%%%%%%%%%%%%%%%%
\begin{figure}[!t]
    \centering
    \includegraphics[width=.9\columnwidth]{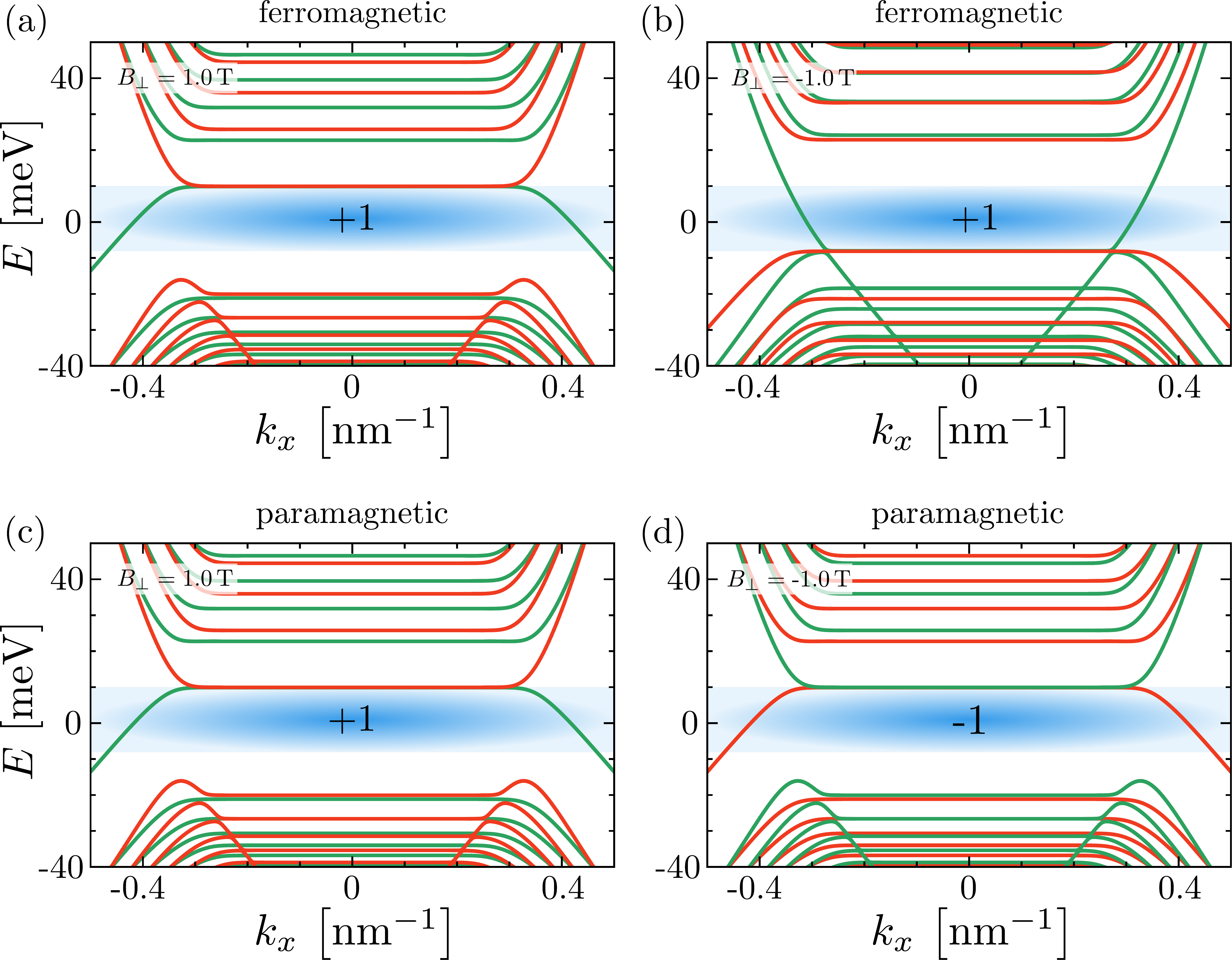}
    \caption{\label{fig:llBHZSymmetryMagQAH}Landau level spectrum of BHZ model in the presence of a (a)-(b) ferromagnetic, or (c)-(d) paramagnetic exchange interaction, where in (a) and (c) $\Bz=1\tesla$, and in (b) and (d) $\Bz=-1\tesla$. The color codes marks the spin up (orange) and spin down (green) block. We employed the following set of parameters: $M=-1\meV$, $B=-685\meV \nm^2$, $D=-600\meV \nm^2$, and $A=365\meV \nm$. (a)-(b) In the ferromagnetic case, we use $g(\Bz)=g_0$ with $g_0=9 \meV$. (c)-(d) In the paramagnetic case, we use $g(\Bz)=g_0 \Bz$ with $g_0=9 \meV / \tesla$.   For this specific set of parameters, the paramagnetic magnetization matches the ferromagnetic one at $\Bz=1\tesla$. The blue regime marks the Dirac mass gap, i.e., the QAH regime. Chern numbers are indicated.}
\end{figure}
%%%%%%%%%%%%%%%%%%%%%%%%%%%%%%%%%%%%%%%%%%

Let us first consider the ferromagnetic case, where the magnetization $g(\Bz)$ follows a hysteresis. For our model, we assume that $g(\Bz)\equiv g_0$ remains constant for a given polarization direction of the magnetic domains. This holds until the external magnetic field exceeds the coercive field $B_c$ of the ferromagnet. Above this threshold, the magnetic domains can flip their polarization direction to align with the external field. At $\Bz=0$, we assume that the system is in the \ac{QAH} phase, i.e., $(M+g_0)(M-g_0)<0$ \cite{Liu08}. This condition guarantees that only one of the two spin blocks is topologically nontrivial. The conductivity in the Dirac mass gap is hence determined solely by the magnetization direction, $\sigma_{xy}=\sgn{g_0} e^2/h$. Applying an external magnetic field, Eqs.~\eqref{eq:qshCond} and \eqref{eq:qahCondPart} show that the Hall conductivity  in the Dirac mass gap (at constant $\mu$) follows in quantized steps the magnetic hysteresis  ($g_0$). This remains valid as long as the orbital contribution in Eq.~\eqref{eq:qahCondPart}, the $B/l_{\Bz}^2$ term, is small compared to the magnetization. More precisely, $\sigma_{xy}$  drops to zero, when the term $B/l_{\Bz}^2$ exceeds the effective Dirac mass, $M\pm g_0$, at $\Bz>B_{\perp,triv}$, where
\begin{align}
    B_{\perp,triv}=\sgn{e \Bz} \frac{\hbar}{e} \max \left( \frac{M \pm g_0}{B} \right) \, .
\end{align}
Ultimately, the orbital contribution drives both spin blocks into the trivial regime. 
The behavior of $\sigma_{xy}$ as a function of $\Bz$ at constant $\mu$ is schematically shown in Fig.~\ref{fig:paraVsFerro}(a), where we assume that $B_c < B_{\perp,triv}$.  Since $\sigma_{xy}$ follows the magnetic hysteresis, the Hall conductivity is an even function of the magnetic field for  $\Bz<B_c$. This represents a violation of the Onsager relation [cf. App.~C]. The peculiar behavior of $\sigma_{xy}$ is encoded in a nonzero spectral asymmetry $\eta_{BHZ}^{}$, which only drops to zero for $\Bz>B_{\perp,triv}$. As we have stated previously, this is a signature of the QAH effect in magnetic field and results as a consequence of the parity anomaly. 

Finally, note that in this scenario, the Dirac mass gap is  defined by
\begin{align} \label{eq:diracGapBHZ}
    \abs{\mu+D/l_{\Bz}^2} \leq \min \abs{M \pm g_0-B/l_{\Bz}^2} \, .
    %\, , \, \abs{M-g_0-\frac{B}{l_{\Bz}^2}}\rb \, .
\end{align}
In comparison to the case of a single Chern insulator, the `minimum' is here required to ensure that the chemical potential is placed within the Dirac mass gap of both spin blocks. When the chemical potential is chosen such that Eq.~\eqref{eq:diracGapBHZ} is fulfilled, a violation of the Onsager relation should be observed in a related experiment.

To analyze the role of the QAH edge states in this (ferromagnetic) case for $\Bz<B_c$, we show exemplary the band structure of a QAH insulator with $g_0>0$  in Figs.~\ref{fig:llBHZSymmetryMagQAH}(a) and~(b) for $\Bz=1\tesla$ and $-1\tesla$, respectively. For the given  system parameters and independent of the magnetic field direction, only the spin down (green) block is in the inverted regime. This is reflected by the existence of only spin down QAH edge states in the Dirac mass gap (blue area). It hence follows that $\sigma_{xy}(-\Bz)=\sigma_{xy}(\Bz)$. Due to the strong PH asymmetry, the appearance of the QAH edge states outside of the Dirac mass gap changes with the magnetic field direction.

%%%%%%%%%%%%%%%%%%%%%%%%%%%%%%%%%%%%%%%%%%%%%%%%%%%%%%%

%%%%%%%%%%%%%%%%%%%%%%%%%%%%%%%%%%%%%%%%%%%%%%%%%%%%%%%

Let us now turn to the (second) paramagnetic case. For simplicity, we assume that $g(\Bz)=g_0 \Bz$,  although one should bare in mind that a  paramagnetic exchange interaction is actually determined by a Brillouin function \cite{Liu08}. Since we are here interested in a qualitative discussion, this approximation allows us to write down analytic results. In comparison to the ferromagnetic case, the system is in the \ac{QSH} phase, i.e., $\sigma_{xy}=0$ at $\Bz=0$ [cf. Fig.~\ref{fig:paraVsFerro}(b)]. Applying now an external magnetic field breaks the symmetry between the two spin blocks, so that the associated Dirac mass gaps close at two different  critical magnetic fields, i.e., $B_{\perp,triv,\spinup} \neq B_{\perp,triv,\spindown}$. Ultimately, the QAH phase is induced when one of the two spin blocks becomes trivial.  From  Eq.~\eqref{eq:qahCondPart}, we see that this is the case if
\begin{align} \label{eq:qahCondMag}
    \lb M+g(\Bz)-\frac{B}{l_{\Bz}^2} \rb \lb M-g(\Bz)-\frac{B}{l_{\Bz}^2} \rb <0 \, .
\end{align}
In comparison to Ref.~\cite{Liu08}, this generalizes the condition for the \ac{QAH} effect to finite external magnetic fields. 
Equation~\eqref{eq:qahCondMag} is fulfilled for $\abs{\Bz}>B_{\perp,QAH}$, where
\begin{align}\label{eq:critMagQAH}
    B_{\perp,QAH}&=\min\left[B_{\perp,triv,\spinup},B_{\perp,triv,\spindown}\right] \\ 
    &=\min \left[ \frac{M}{\sgn{e\Bz}B e /\hbar \mp g_0}\right] \, .
    %\, , \, \frac{M}{\sgn{e\Bz}B e /\hbar+g_0} \rb \, .
\end{align}
In this regime, the Hall conductivity is given by $\sigma_{xy}=\sgn{g_0}\sgn{e \Bz} e^2/h$ (for a constant  chemical potential within the Dirac mass gap). The $\sgn{e \Bz}$ dependence results from the fact that, depending on the magnetic field direction, either $\sigma_{xy,\spinup}^I$ or $\sigma_{xy,\spindown}^I$ is nonzero. Paramagnetic topological insulator therefore do not violate the Onsager relation 
even though each spin block can exhibit the parity anomaly. We can understand this effect also in terms of band structure calculations. As an example, we show in Figs.~\ref{fig:llBHZSymmetryMagQAH}(c) and~(d) the band structures for $\Bz=1\tesla$ and $-1\tesla$, respectively.  Depending on $\sgn{e \Bz}$, the QAH edge states of either the spin up (orange) or down (green) block traverse the Dirac mass gap.  Recall that their chirality is determined by their intrinsic Chern number and not by the magnetic field. Although the Onsager relation is therefore not violated, the survival of the QAH edge states in the Dirac mass gap is still apparent.

When the magnetic field is increased further, the system becomes ultimately a trivial insulator  when the  Dirac mass gap of the remaining, second spin block is closed at [cf. Eq.~\eqref{eq:qahCondPart}]
\begin{align}\label{eq:critMagTriv}
    B_{\perp,triv}&=\max\left[B_{\perp,triv,\spinup},B_{\perp,triv,\spindown}\right] \\ 
    &=\max \lb \frac{M}{\sgn{e\Bz}B e /\hbar \mp g_0} \rb \, .
    %\, , \, \frac{M}{\sgn{e\Bz}B e /\hbar+g_0} \rb \,.
\end{align}
Above this threshold, the Hall conductivity drops to zero \footnote{The statement holds provided that the non-relativistic mass in Eq.~\eqref{eq:critMagTriv} dominates in the large $\Bz$-limit, $|B/l_{\Bz}^2|>|g_0|$.}. As sketched in Fig.~\ref{fig:paraVsFerro}(b), the Hall conductivity (at constant $\mu$) evolves therefore from 0 to $\pm e^2 /h$, and again to 0 with increasing $\abs{\Bz}$. This so-called reentrant behavior of $\sigma_{xy}$ \cite{Beugeling12} is hence encoded in the spectral asymmetry and can be interpreted as a representative of the parity anomaly.

%%%%%%%%%%%%%%%%%%%%%%%%%%%%%%%%%%%%%%%%%%%
%%%%		SUMMARY & OUTLOOK
%%%%%%%%%%%%%%%%%%%%%%%%%%%%%%%%%%%%%%%%%%%
\section{Summary and Outlook} \label{sec:outlook}
In conclusion, we have derived the relation between the spectral asymmetry (parity anomaly) and the QAH effect in magnetic fields in the presence of a broken particle-hole symmetry. 
We showed that, in contrast to the parity breaking mass terms, a PH asymmetry acts like a magnetic-field-dependent chemical potential. 
In addition, we showed that the strength of the PH asymmetry strongly affects the hybridization of coexisting QH and QAH edge states in magnetic fields. 
Finally, we predicted experimental signatures that are connected to the parity anomaly in para- and ferromagnetic TIs. 
Interesting future directions include analyzing the temperature dependence of the spectral asymmetry, as well as extending the scope of this topological quantity to other topological materials, such as 3D TIs or Weyl semimetals, which are also governed by  Dirac-like operators.

\begin{acknowledgments}
We thank B.~Scharf, R.~Meyer, B.~Trauzettel, B.~A.~Bernevig, F.~Wilczek, C.~Br\"une, J.~S.~Hofmann, J.~Erdmenger, C.~Morais Smith, and W.~Beugeling for useful discussions.

We acknowledge financial support through the Deutsche Forschungsgemeinschaft (DFG, German Research Foundation), Project-id No. 258499086--SFB 1170 ``ToCoTronics'', the ENB Graduate School on Topological Insulators and through the W\"urzburg-Dresden Cluster of Excellence on Complexity and Topology in Quantum Matter--ct.qmat (EXC 2147, Project-id No.  39085490).
\end{acknowledgments}

 %%BIBLIOGRAPHY
\bibliographystyle{apsrev4-1}

%%%%%%%%%%%%%%%%%%%%%%%%%%%%%%%%%%%%%%%%%%%%%%%%%%%%%%%%%%%%%%%%%%%%%%%%%%%%%%%%%%%%%%%
%%%%%%%%%%%%%%%%%%%%%%%%%%%%%%%%%%%%%%%%%%%%%%%%%%%%%%%%%%%%%%%%%%%%%%%%%%%%%%%%%%%%%%%
%%%%%%%%%%%%%%%%%%%%%%%%%%%%%%%%%%%%%%%%%%%%%%%%%%%%%%%%%%%%%%%%%%%%%%%%%%%%%%%%%%%%%%%
%%%%%%%%%%%%%%			APPENDICES
%%%%%%%%%%%%%%%%%%%%%%%%%%%%%%%%%%%%%%%%%%%%%%%%%%%%%%%%%%%%%%%%%%%%%%%%%%%%%%%%%%%%%%%
%%%%%%%%%%%%%%%%%%%%%%%%%%%%%%%%%%%%%%%%%%%%%%%%%%%%%%%%%%%%%%%%%%%%%%%%%%%%%%%%%%%%%%%
%%%%%%%%%%%%%%%%%%%%%%%%%%%%%%%%%%%%%%%%%%%%%%%%%%%%%%%%%%%%%%%%%%%%%%%%%%%%%%%%%%%%%%%

\appendix
%%%%%%%%%%%%%%%%%%%%%%%%%%%%%%%%%%%%%%%%%%%%%%%%%%%
%%%%%%%             EIGENSTATES
%%%%%%%%%%%%%%%%%%%%%%%%%%%%%%%%%%%%%%%%%%%%%%%%%%%
\section{Wave functions of BHZ model} \label{app:Wavefunctions}

We consider a Chern insulator given by Eq.~\eqref{eq:chernInsulator}
(`spin up' block of \ac{BHZ} model) and derive the corresponding wave functions and field operators in the absence, as well as in the presence of an external out-of-plane  magnetic field. As discussed in the main text, solutions for the spin down block can be obtained by replacing $M \rightarrow -M$ and $B\rightarrow -B$.  To keep the notation neat, we drop the spin index in the following.

Appropriate solutions of the Schr\"odinger equation in the absence of an external magnetic field are given by:
\begin{align} \label{eqA:wavefunctionNoBz}
    u_{k_x,k_y}( \xvec ) &=c_1
    \begin{pmatrix}
    M-B k^2+\epsilon (k) \\
    A k_-
    \end{pmatrix}
    \ee^{\ii \kvec \xvec } \, ,\\
    %%%%%%%%%%%%%%%%%%%%
    v_{k_x,k_y}( \xvec ) &=c_2
    \begin{pmatrix}
    M-B k^2-\epsilon(k) \\
    A k_-
    \end{pmatrix}
    \ee^{\ii \kvec \xvec } \, ,
\end{align}
where $\epsilon(k) = \sqrt{A^2 k^2+\lb M- B k^2 \rb^2}$, and $c_{1,2}$ are normalization constants. Here, $u_{k_x,k_y}(\xvec)$ is a solution of the conduction band and $v_{k_x,k_y}(\xvec)$ of the valence band corresponding to $E^+_{\spinup} (\kvec)$ and $E^-_{\spinup} (\kvec)$ [Eq.~\eqref{eq:spectrumCI})], respectively. The time-independent field operator $\psi(\xvec)$ can be expanded in the basis formed by $u_{k_x,k_y}(\xvec)$ and $v_{k_x,k_y}(\xvec)$:
\begin{align} \label{eq:fieldOp}
	\Psi \lb \xvec \rb =\sum_{k_x,k_y}\left[ b_{k_x,k_y}u_{k_x,k_y}(\xvec)+ d_{k_x,k_y}^\dagger v_{k_x,k_y}(\xvec) \right] \, .
\end{align}

Subjecting the system to an external magnetic field gives rise to the formation of \acp{LL}. Appropriate solutions of the Schr\"odinger equation in the Landau gauge for positive magnetic fields are determined by
\begin{align} \label{eqA:wavefunctionPosBz}
      \ket{\psi_{n\neq 0,k_x}^\pm}=
    %%%%%%%%%%%%%
        c_n
        \begin{pmatrix}
            \lb M-\frac{2Bn-D}{l_{\Bz}^2} \pm \epsilon_n \rb \ket{n,k_x} \\
            \sqrt{\frac{2 A n}{l_{\Bz}^2}} \ket{n-1,k_x}
        \end{pmatrix}
\end{align}
and
\begin{align}
    %%%%%%%%%%%%
    \ket{\psi_{n=0,k_x}^\pm}=
        \begin{pmatrix}
            \ket{0,k_x} \\
            0
        \end{pmatrix} \, ,
\end{align}
where
\begin{align}
\epsilon_n=\sqrt{\frac{2 n A^2}{l_{\Bz}^2}+\lb M - \frac{2 B n+\sgn{\Bz}D}{l_{\Bz}^2} \rb^2} \, .
\end{align}
The superscript marks whether a state is part of the valence ($-$) or the conduction ($+$) band. Solutions for negative magnetic fields are given by
\begin{align} \label{eqA:wavefunctionNegBz}
      \ket{\psi_{n\neq0,k_x}^\pm}=
    %%%%%%%%%%%%%
        c_n \!\!
        \begin{pmatrix}
            \lb M-\!\frac{2Bn+D}{l_{\Bz}^2} \pm \epsilon_n \rb \ket{n-1,k_x} \\
            -\sqrt{\frac{2 A n}{l_{\Bz}^2}} \ket{n,k_x}
        \end{pmatrix}
\end{align}
and
\begin{align}
    \ket{\psi_{n=0,k_x}^\pm}=
    %%%%%%%%%%%%
        \begin{pmatrix}
            0 \\
            \ket{0,k_x}
        \end{pmatrix} \, ,
\end{align}
and $c_n$ defines the appropriate normalization constant. 
We can therefore write the field operator in magnetic fields in terms of the \ac{LL} spinors, 
\begin{align} \label{eqA:fieldOpMag}
	\Psi \lb \xvec \rb =\sum_{n,k_x}b_{n,k_x}u_{n,k_x}(\xvec)+\sum_{n,k_x} d_{n,k_x}^\dagger v_{n,k_x}(\xvec) \, ,
\end{align}
where $u_{n,k_x}(\xvec)=\ee^{\ii k_x x} \braket{y|\psi_{n,k_x}^+}$ labels again a solution of the conduction band and $v_{n,k_x}(\xvec)=\ee^{\ii k_x x} \braket{y|\psi_{n,k_x}^-}$ a solution of the valence band. The $n=0$ \ac{LL} plays a special role in this context. Since it is either part of the valence band or the conduction band, it can contribute either only to the first or to second sum in Eq.~\eqref{eqA:fieldOpMag}.  More precisely, the first sum in Eq.~\eqref{eqA:fieldOpMag} runs from $n=1$ to $\infty$ and the second sum from $n=0$ to $\infty$, if the $n=0$ \ac{LL} belongs to the valence band. The situation is in reverse if the $n=0$ \ac{LL} is part of the conduction band. The situation is similar to a purely relativistic Dirac fermion in (2+1)D \cite{Abouelsaood85,Boyanovsky86,Haldane88}.

%%%%%%%%%%%%%%%%%%%%%%%%%%%%%%%%%%%%%%%%%%%%%%%%%%%
%%%%%%%             SYMMETRIES
%%%%%%%%%%%%%%%%%%%%%%%%%%%%%%%%%%%%%%%%%%%%%%%%%%%

\section{Symmetries of Chern insulator and BHZ model} \label{app:symmetries}

In this section, we review the symmetries of a general Hamiltonian in two space dimensions in the presence of an external vector potential $\mathbf{A}(\xvec)$. The Hamiltonian in second quantization reads
 \begin{align}\label{eqA:secondQuantizedHam}
    \mathcal{H}=\myInt{\xvec} \Psi^\dagger (\xvec) H\left[-\ii \boldsymbol{\nabla};\mathbf{A}(\xvec)\right] \Psi(\xvec) \, .
\end{align}
 We are in particular interested in the following discrete symmetries: time-reversal ($\mathcal{T}$), parity ($\mathcal{P}$), particle-hole ($\mathcal{C}$), and chiral ($\mathcal{S}$) symmetry. The operators $\mathcal{T}$ and $\mathcal{S}$ are antiunitary, while $\mathcal{P}$ and $\mathcal{C}$ are unitary. The field operator transforms in the following way \cite{Chiu16}:
\begin{align}
    \mathcal{T} \Psi(\xvec) \mathcal{T}^{-1} &= U_T \Psi(\xvec) \label{eqA:trsField} \, , \\
    \mathcal{C} \Psi(\xvec) \mathcal{C}^{-1} &= U_C^* \Psi^\dagger(\xvec) \label{eqA:phsField} \, , \\
    \mathcal{S} \Psi(\xvec) \mathcal{S}^{-1} &= U_S \Psi^\dagger(\xvec) \label{eqA:chiralField} \, , \\
    \mathcal{P}_x \Psi(x,y) \mathcal{P}_x^{-1} &= U_P \Psi(-x,y) \label{eqA:parityField} \, ,
\end{align}
where $U_T$, $U_P$, $U_C$, and $U_S$ are unitary matrices. Note that in (2+1)D a parity transformation can be defined either as $\mathcal{P}_x:(x,y)\rightarrow (-x,y)$ or as $\mathcal{P}_y:(x,y)\rightarrow (x,-y)$. This is in contrast to a system defined in a (3+1)D space, where a parity transformation flips the sign of all spatial components, i.e., $\mathcal{P}:(x,y,z)\rightarrow(-x,-y,-z)$. Since $\mathbf{A}(\xvec)$ denotes an \textit{external} vector potential, it is unaffected by any non-spatial symmetry operation. Under a spatial symmetry, like parity, it transform as
\begin{align}\label{eqA:transGauge}
    \mathcal{P}_x \mathbf{A}(x,y) \mathcal{P}_x^{-1}=\mathbf{A}(-x,y) \, .
\end{align}
The direction of an external magnetic field does therefore not change under parity. In comparison, if the physical source of the vector potential is an integral part of our closed system (as assumed in (2+1)D \ac{QED}), it would transform under both spatial and non-spatial symmetry operations \cite{Deser82}. In case of the parity transformation, this would imply that $\mathcal{P}_x \mathbf{A}(x,y) \mathcal{P}_x^{-1}=-\mathbf{A}(-x,y)$. Opposed to our system, a magnetic field is therefore considered as a pseudoscalar in (2+1)D \ac{QED} \cite{Boyanovsky86}.

An arbitrary linear and unitary, or antilinear and antiunitary operator $\mathcal{O}$ describes a symmetry if $\commutator{ \mathcal{H} , \mathcal{O} }=0$~\cite{Chiu16}.  We derive the symmetry conditions for our non-interacting, single particle Hamiltonian $H$ by inserting Eqs.~\eqref{eqA:trsField}~--~\eqref{eqA:parityField} and Eq.~\eqref{eqA:transGauge} into Eq.~\eqref{eqA:secondQuantizedHam}. The relations corresponding to the four different discrete symmetries are given by
\begin{align}
U_T^\dagger H^*\left[\kvec;\mathbf{A}(\xvec)\right]U_T &=H\left[-\kvec;\mathbf{A}(\xvec)\right] \label{eqA:unitaryTrafoTR} \, ,\\
U_C^\dagger H^*\left[\kvec;\mathbf{A}(\xvec)\right]U_C &=-H\left[-\kvec;\mathbf{A}(\xvec)\right] \label{eqA:unitaryTrafoPH} \, ,\\
U_S^\dagger H\left[\kvec;\mathbf{A}(\xvec)\right]U_S &=-H\left[\kvec;\mathbf{A}(\xvec)\right] \label{eqA:unitaryTrafoChiral} \, ,\\
U_P^\dagger H\left[\kvec ;\mathbf{A}(\xvec)\right]U_P &=H\left[\kvec^\prime;\mathbf{A}(\xvec^\prime)\right] \, , \label{eqA:unitaryTrafoParity}
\end{align}
where $\kvec^\prime=(-k_x,k_y)$ and $\xvec^\prime = (-x,y)$. 
Here, we used a Fourier transformation of $\Psi(\xvec)$ to work in momentum space. This choice is convenient in the absence of an external vector potential because  translational invariance in both spatial directions implies that the Hamiltonian is diagonal in momentum space. This is not the case in the presence of an external field. Nonetheless, we derived the symmetry conditions in momentum space to permit comparability between both cases.

In the following, we consider two examples for which the first quantized Hamiltonian $H$ describes either a Chern insulator or the full \ac{BHZ} model. Moreover, we focus on an external magnetic field with $\mathbf{A}(\xvec)=-y\Bz \mathbf{e}_x$.

%%%%%%%%%%%%%   SUB: CHERN INSULATOR
\subsection{Chern insulator}

A Chern insulator exhibits only all of the discussed discrete symmetries if $M=B=D=0$ and if no external field is applied. In this case, we find that
\begin{align}
    U_T=-\ii \sigma_y \, , \quad U_C=\sigma_x \, ,  \quad U_S=\sigma_z \, , \,\,\,\, U_P=\sigma_y \, .
\end{align}
Introducing either the Dirac mass $M$ or the non-relativistic mass $B$ breaks \ac{TR}, parity, and chiral symmetry. A non-zero $D$ parameter breaks the \ac{PH} and chiral symmetry. What is the effect of an external magnetic field? It breaks all symmetries except for the chiral symmetry. This is due to the fact that the $n=0$ \ac{LL} is in this case exactly at zero energy and is therefore its own partner under a chiral transformation.

%%%%%%%%%%%%%   SUB:  BHZ MODEL
\subsection{BHZ model}
In case of the \ac{BHZ} model, the unitary matrices are given by
\begin{align}
    &U_T=-\ii \tau_y \otimes \sigma_0 \, , \quad
    U_C=\tau_0 \otimes \sigma_x \, ,  \quad \nonumber \\
    &U_S=\tau_x \otimes \sigma_x \, ,  \quad
    U_P=\tau_y \otimes \sigma_0\, .
\end{align}
Let us again consider the effect of each parameter separately. In contrast to a Chern insulator, neither $M$ nor $B$ break any discrete symmetry. This is possible because $\mathcal{T}$, $\mathcal{S}$, and $\mathcal{P}$ involve both spin blocks. The Zeeman term $g$ takes the role of $M$ and $B$ in a single Chern insulator and breaks \ac{TR}, parity, and chiral symmetry. The $D$ parameter breaks \ac{PH} and chiral symmetry.  Again, an external magnetic field breaks all symmetries except for the chiral symmetry.

\subsection{Parity symmetry of charge}
We briefly want to comment on the parity symmetry of charge in our system compared to (2+1)D \ac{QED}. In the latter, charge is considered to be a pseudoscalar which means that it is odd under a parity transformation \cite{Abouelsaood85,Boyanovsky86,Schakel91}. This is because $\rho(\Bz)=e^2 \sgn{M} \Bz /2h$, where $\Bz$ is odd under parity \footnote{In $\sgn{M}$, $M$ is simply a parameter. It is therefore even under parity. This should not be confused with the mass operator $M\sigma_z$ which is odd under parity due to the unitary part of transformation.}. This is in contrast to our system, where the magnetic field and, hence,  the charge are both conventional scalars [cf. with discussion below Eq.~\eqref{eqA:transGauge}].

%%%%%%%%%%%%%%%%%%%%%%%%%%%%%%%%%%%
%% ONSAGER RELATION
%%%%%%%%%%%%%%%%%%%%%%%%%%%%%%%%%%%
\section{Onsager relation} \label{app:Onsager}
In a conventional QH phase, the Onsager relation imply that $\sigma_{xy}(\Bz)=-\sigma_{xy}(-\Bz)$ \cite{AshcroftBook}. When we refer to a violation of the Onsager relation, we want to highlight that, in contrast to a conventional QH phase, the Hall conductivity of a QAH insulator is an even function of the magnetic field if the chemical potential is placed in the Dirac mass gap, i.e.,  $\sigma_{xy}(\Bz)=\sigma_{xy}(-\Bz)$. In this regime, the difference to the QH phase arises because the Hall conductivity is  determined by the intrinsic topology of the QAH insulator [Eq.~\eqref{eq:intrinsicHall}], rather than by the magnetic field. Hence, the Hall conductivity only switches its sign if we flip both the sign of the intrinsic Chern number, $(M,B)\rightarrow(-M,-B)$, as well as the sign of the external magnetic field. A QAH insulator is therefore characterized by
\begin{align}
    \sigma_{xy}(M,B,\Bz)=-\sigma_{xy}(-M,-B,-\Bz) \, .
\end{align}

%%%%%%%%%%%%%%%%%%%%%%%%%%%%%%%%%%%
%%% SOME INTERMEDIATE STEPS
%%%%%%%%%%%%%%%%%%%%%%%%%%%%%%%%%%%
\section{Spectral asymmetry} \label{app:spectralAsym}
%%%%%%%%%%%%%%%%%%%%%%%%%%%%%%%%%%%%%%%%%%%%%%%%%
\subsection{No external magnetic field} 
In writing Eq.~\eqref{eq:etaIntegralForm}, we omitted terms in the exponent which are $\mathcal{O}(k^{-2})$ since they vanish in the limit $\kappa \rightarrow 0^+$ at the end of the calculation. Let us proof this assertion. We start with
\begin{multline}
    \etaVacR=\frac{S}{(2\pi)^2} \ee^{-\kappa \, \sgn{B} \lb \frac{A^2}{2B}-M \rb } 
    \left[\ee^{\kappa E_z}\int_{\mathbb{R}^2_{}} d \kvec \, \ee^{-\kappa B_- k^2} \right.  \\
     \left. \times \ee^{-\kappa \mathcal{O}(k^{-2})}
    -\ee^{-\kappa E_z}\int_{\mathbb{R}^2_{}} d \kvec \, \ee^{-\kappa B_+ k^2}\ee^{-\kappa \mathcal{O}(k^{-2})} \right] \, ,
\end{multline}
where $B_\pm=\abs{B}\pm D$.
We have to show that all higher order terms (higher than zeroth order) in $\exp \left[-\kappa \mathcal{O}(k^{-2})\right]$ are on the order $\mathcal{O}(\kappa)$ and, therefore, vanish in the limit $\kappa \rightarrow 0^+$. This implies that we have to evaluate integrals of the type
\begin{align}
    \int_{\mathbb{R}^2_{}} d \kvec  \, \ee^{-\kappa c k^2} \ee^{-\kappa (k^{-2})^m} \, ,
\end{align}
with $m=1,2,3,\ldots$ and $c>0$ is a positive constant. Let us consider the case $m=1$ explicitly.
\begin{align}
    \int_{\mathbb{R}^2_{}} d \kvec  \, \ee^{-\kappa c k^2} \ee^{-\kappa k^{-2}}=\pi \int_0^\infty ds \, \ee^{-\kappa c s } \,  \ee^{-\kappa / s}  \nonumber\\
    =2 \sqrt{c^{-1}} K_1\lb 2\sqrt{c}\kappa \rb 
    =\frac{1}{c\kappa}+ \mathcal{O}(\kappa) \, ,
    %=\frac{1}{c\kappa} -\kappa \left[ 1-2 \gamma-\log \lb c \kappa^2\rb \right]+ \mathcal{O}(\kappa^3) \, , 
\end{align}
where we employed polar coordinates, as well as the substitution $s=k^2$, and $K_1$ is the modified Bessel function of the second kind. %$\gamma\approx 0.577$ is Euler's constant 
We have therefore proven our assertion for $m=1$. By analogously evaluating integrals with $m>1$, one verifies that the assertion holds for all $m$. 

%%%%%%%%%%%%%%%%%%%%%%%%%%%%%%%%%%%%%%%%%%%%%%%
\subsection{Finite magnetic fields}
To arrive at Eq.~\eqref{eq:spectralAsymSDFirst}, we used the following Taylor expansion for large $n$
\begin{align}
    E_{n\neq0}^{\pm}&=-\frac{\sgn{e \Bz} B + 2 n D}{l_{\Bz}^2}  \pm \frac{2 n \abs{B}}{l_{\Bz}^2} \sqrt{1+x} \\
    &=-\frac{\sgn{e \Bz} B + 2 n D}{l_{\Bz}^2}  \pm \frac{2 n \abs{B}}{l_{\Bz}^2}\left[ 1+\frac{x}{2}+\mathcal{O}(x^2)\right] \, 
\end{align}
where
\begin{multline}
x=\frac{1}{4n^2 B^2 l_{\Bz}^{-4}} \left[M^2+2 n l_{\Bz}^{-2} \lb A^2 -2  M B \rb \right. \\
\left. -2 \,  \sgn{e \Bz}  D l_{\Bz}^{-2} \left( M -2 n B  l_{\Bz}^{-2}  \right) + D^2 l_{\Bz}^{-4} \right] \, .
\end{multline}
We can reinsert $x$ to see that terms with $n \mathcal{O}(x^2)$ are of the order  $\mathcal{O}(n^{-1})$. We can recast the equation in the following way
\begin{multline} \label{eqA:llApprox}
    E_{n\neq0}^{\pm}=
    - \frac{ 2 n }{l_{\Bz}^2} \left( D \mp  \abs{B} \right)
    -\frac{\sgn{e \Bz} B }{l_{\Bz}^2} \\
    \pm \frac{\lb A^2 -2  M B \rb}{2 \abs{B}}
    \pm \frac{\sgn{e \Bz} D B }{l_{\Bz}^{2} \abs{B}}+\mathcal{O}(n^{-1}) \, .
\end{multline}
In order now to evaluate the infinite sums in Eq.~\eqref{eq:spectralAsymSDFirst} using \eqref{eqA:llApprox}, we have to consider sums of the type
\begin{align}
   \sum_{n=1}^\infty  \ee^{-\kappa c n } \ee^{-\kappa \mathcal{O}\lb n^{-1} \rb} \, .
\end{align}
Analogously to the $\Bz=0$ case from above, we can show that it is sufficient to keep only the lowest order term in $\exp \left[ \mathcal{O}(n^{-1}) \right]$. All higher order terms vanish. To this end, let us study the following infinite series 
\begin{align}
   \sum_{n=1}^\infty \frac{\kappa}{n}  \ee^{-\kappa c n } =-\kappa \log \lb 1- \ee^{-c \kappa } \rb = \mathcal{O} (\kappa) \, . 
\end{align}
As the sum is clearly on the order $\mathcal{O}(\kappa)$, it vanishes in the limit $\kappa \rightarrow 0^+$. We can repeat these steps for all higher order terms, where we use the polygarithm function. This implies that only the lowest order term with $\exp \left[\mathcal{O}(n^{-1}) \right] \approx 1$ contributes to the final result. This term is a geometric series and can be easily evaluated.

\end{document}